\documentclass[pdflatex,sn-mathphys-num,iicol]{sn-jnl}

\usepackage[utf8]{inputenc}
\usepackage[T1]{fontenc}

\usepackage{graphicx}%
\usepackage{multirow}%
\usepackage{amsmath,amssymb,amsfonts}%
\usepackage{amsthm}%
\usepackage{mathrsfs}%
\usepackage[title]{appendix}%
\usepackage{xcolor}%
\usepackage{textcomp}%
\usepackage{manyfoot}%
\usepackage{booktabs}%
\usepackage{algorithm}%
\usepackage{algorithmicx}%
\usepackage{algpseudocode}%
\usepackage{listings}%

\usepackage{amsmath,tabstackengine}
\usepackage{amsfonts}
\usepackage{tabularx}
\usepackage{framed}
\usepackage{times}
\usepackage[utf8]{inputenc}
\usepackage{pgfplots}
\usepackage{pgfplotstable}
\usepackage{tikz}
\usetikzlibrary{decorations.markings, arrows.meta, positioning, shapes.geometric}
\pgfplotsset{compat=1.15}
\usepackage{colortbl}
\usepackage{lipsum} 
\usepackage{mathtools} 
\usepackage{array}
\usepackage{makecell}
\usepackage{booktabs}
\usepackage{graphicx}
\usepackage{multirow}
\usepackage{color}
\usepackage{braket}
\usepackage{dcolumn}
\usepackage{siunitx}
\usepackage{bm,url}

\begin{document}

\title[Article Title]{Non-minimally coupled Weyl connection gravity in the Solar System and at the Galactic Center}

\author[1,2,3]{\fnm{Mohsen } \sur{Khodadi}}\email{m.khodadi@du.ac.ir}

\author*[4,5,6]{\fnm{Margarida} \sur{Lima}}\email{margarida.a.lima@tecnico.ulisboa.pt}
\equalcont{These authors contributed equally to this work.}

\author[6,7]{\fnm{Cl\'{a}udio} \sur{Gomes}}\email{claudio.fv.gomes@uac.pt}
\equalcont{These authors contributed equally to this work.}

\affil*[1]{\orgdiv{School of Physics}, \orgname{Institute for Research in Fundamental Sciences (IPM)}, \orgaddress{\postcode{19395-5531}, \city{Tehran}, \country{Iran}}}

\affil[2]{\orgdiv{School of Physics}, \orgname{Damghan University}, \orgaddress{\city{Damghan}, \postcode{3671645667}, \country{Iran}}}

\affil[3]{\orgdiv{Center for Theoretical Physics}, \orgname{Khazar University}, \orgaddress{\street{41 Mehseti Str.}, \city{Baku}, \postcode{AZ1096}, \country{Azerbaijan}}}

\affil[4]{\orgdiv{Departamento de F\'{i}sica }, \orgname{ Instituto Superior T\'{e}cnico, Universidade de Lisboa}, \orgaddress{\street{ Av. Rovisco Pais}, \city{ Lisboa}, \postcode{ 1049-001}, \country{ Portugal}}}

\affil[5]{\orgdiv{Centro de An\'{a}lise Matem\'{a}tica, Geometria e Sistemas Din\^{a}micos}, \orgname{Departamento de Matem\'{a}tica, Instituto Superior T\'{e}cnico, Universidade de Lisboa}, \orgaddress{\street{Av. Rovisco Pais }, \city{ Lisbon}, \postcode{ 1049-001}, \country{ Portugal}}}

\affil[6]{\orgdiv{ OKEANOS--Instituto de Investiga\c{c}\~{a}o em Ci\^{e}ncias do Mar}, \orgname{Universidade dos A\c{c}ores}, \orgaddress{\street{ Rua Prof. Doutor Frederico Machado, 4}, \city{Horta }, \postcode{9900-140  }, \country{ Portugal}}}

\affil[7]{\orgdiv{Centro de F\'{i}sica das Universidades do Minho e do Porto}, \orgaddress{\street{ Rua do Campo Alegre s/n}, \city{Porto}, \postcode{4169-007}, \country{ Portugal}}}


\abstract{We explore the phenomenological viability of non-minimally coupled Weyl connection gravity by confronting its static, spherically symmetric black hole solutions with classical weak-field Solar System tests and stellar-orbit observations near Sgr A*. In this geometric framework, non-metricity is encoded via a Weyl vector field, giving rise to two distinct families of Schwarzschild-like vacuum solutions characterized by a free parameter \(\omega\) with dimensions of length. We compute the corrections to four classical observables---gravitational redshift, Mercury's perihelion advance, light deflection, and radar echo delay---and derive stringent lower bounds on \(\omega\) using current observational data. For Solution I (purely radial Weyl vector), the leading metric corrections scale as \(1/\omega\), yielding bounds as strong as \(\omega \gtrsim 10^{30}\) m from perihelion precession. For Solution II (time-radial Weyl vector), the corrections scale as \(1/\omega^2\), resulting in weaker constraints, with \(\omega \gtrsim 10^{20}\) m from the same test. This marked difference arises from the distinct behavior of the linear corrections in each solution: while Solution I exhibits an unsuppressed linear term \(2r/\omega\) that dominates in the Solar System regime, Solution II features a linear term suppressed by an additional factor of \(M/\omega\), making the quadratic term \(-r^2/4\omega^2\) dominant throughout.

We then analyze stellar orbits near the Galactic Center, finding that current observations of the S2 star provide complementary constraints on both solutions. For Solution I, the S2 periastron precession yields \(\omega \gtrsim 10^{21}\) m, while the S2 gravitational redshift provides the stronger constraint \(\omega \gtrsim 10^{24}\) m. For Solution II, the corresponding bounds are \(\omega \gtrsim 10^{17}\) m from periastron precession and \(\omega \gtrsim 10^{22}\) m from gravitational redshift. While Solar System tests probe the weak-field regime, these stellar-orbit observations extend the analysis to a substantially stronger gravitational environment. The stellar-orbit analysis reveals a qualitative distinction: Solution I produces a retrograde correction to the periastron precession, while Solution II produces a prograde one, mirroring the behavior identified in the Solar System. Finally, despite both solutions sharing the same quadratic asymptotic structure, only Solution II can accommodate an effective cosmological constant of the same order as the observed value, \(\Lambda_{\mathrm{obs}} \sim 10^{-52}\) m\(^{-2}\), while Solution I is constrained to produce a negligibly small effective \(\Lambda\).
}




\maketitle

\section{Introduction}	\label{sec:intro}

General Relativity (GR) is a cornerstone in understanding the nature of gravity over a wide range of scales, having successfully passed observational tests from the Solar System to astrophysical scales~\cite{Will:2014kxa}. Nevertheless, several conundrums remain, including the need for large-scale dark components?namely dark matter and dark energy?the cosmological constant problem, the existence of singularities, and the lack of a consistent quantum extension of gravity. These open questions have motivated a plethora of alternative scenarios in the literature, ranging from the inclusion of higher-order curvature invariants, the introduction of additional scalar, vector, or tensor fields, to the use of spacetime torsion and non-metricity~\cite{Capozziello:2011,Odintsov:2011,Odintsov:2017,BeltranJimenez:2017tkd,BeltranJimenez:2019esp,Heisenberg:2018vsk}. 

One such class of models relies on a non-minimal matter-curvature coupling, characterized by two generic functions of the scalar curvature~\cite{nmc}. This framework naturally leads to a non-conservation of the energy-momentum tensor, providing a mechanism to mimic dark matter and late-time cosmic acceleration without introducing exotic components~\cite{Bertolami:2009ic,Bertolami:2010cw,Bertolami:2014hpa}. Remarkably, these models have been shown to pass Solar System tests~\cite{Bertolami:2013qaa,Bertolami:2014gva,March:2016xav,March:2021mqu} and have been extensively studied in various astrophysical and cosmological contexts~\cite{Harko:2011kv,Capozziello:2011wg}.

A natural extension of this framework consists in incorporating Weyl's connection~\cite{Weyl:1918} into the non-minimally coupled model, leading to the so-called non-minimally coupled Weyl connection gravity~\cite{Gomes:2018sbf}. This theory is particularly appealing because it yields second-order metric field equations due to the presence of the Weyl vector field, thereby avoiding the Ostrogradsky instabilities that often plague higher-derivative theories. The model possesses a viable minimum energy space~\cite{Gomes:2018sbf}, is stable under cosmological perturbations~\cite{Baptista:2019,Baptista:2020,Lima:2025ztp}, and its weak-field limit is well-behaved, recovering GR behavior for specific physically motivated ans\"atze for the Weyl vector~\cite{Gomes:2025oxz}. 

Non-metricity, the geometric property encoded by the Weyl connection, has recently attracted considerable attention as it leads to observable signatures that can be tested against local, astrophysical, and cosmological experiments~\cite{Cheng:1988zx,Neeman:1996zcr,Alvarez:2016qfa,Harko:2018gxr,Kouniatalis:2024gnr,Sudharani:2025lxo,Khodadi:2026zoi,Khodadi:2025upl}. In particular, non-metricity-based theories have been explored as potential explanations for dark matter and dark energy~\cite{Bak:2022nrv,Tarasov:2023xpv}. Moreover, the non-metricity raises discussion on the use of geodesics, which gives the "shortest path", or autoparalells, which gives the "straightest path", as the proper description for a particle motion. This discussion has paved interesting insights into the nature of spacetime itself \cite{BarroseSa:1997,Pireaux:2004a,Pireaux:2004b,vandeVenn:2026cjb}. Moreover, another degeneracy lifting comparatively to GR concerns the use of proper time or the affine parameter as the proper clock, which may lead to very different outcomes in atomic clocks and was one of Einstein's objection to Weyl's original theory, despite recent clarifications on the subject \cite{Hobson:2020,Hobson:2021}. In fact, one has to carefully address these issues, particularly when considering modified gravity theories and when there are other properties besides non-metricity, such as a metric field and/or torsion.

Black hole solutions of the vacuum gravitational field equations serve as powerful probes of both the physical viability and the mathematical consistency of any gravitational theory. The observational confirmation of their existence came with the first images of the plasma surrounding the supermassive black hole at the center of galaxy M87, obtained by the Event Horizon Telescope (EHT) collaboration~\cite{EventHorizonTelescope:2019dse}, followed by the imaging of Sagittarius A* (Sgr A*), the black hole at the center of the Milky Way~\cite{EventHorizonTelescope:2022wkp}. These landmark observations not only confirmed key predictions of GR but also opened new avenues for investigating compact astrophysical objects and testing gravitational theories in the strong-field regime~\cite{Khodadi:2020jij,Khodadi:2021gbc,Khodadi:2020gns,Vagnozzi:2022moj,Afrin:2022ztr,Khodadi:2022pqh,Khodadi:2022ulo,Khodadi:2024ubi,Liu:2025wwq}. Black hole solutions, including Schwarzschild- and Reissner-Nordstr\"om-like configurations, have been derived~\cite{Lima:2024cys}, and their shadows have been compared with EHT data~\cite{Gomes:2026qgm}, revealing compatibility for a wide range of the theory's parameters.

Similarly to Weyl squared gravity, whose black hole solutions were derived in Ref.~\cite{Yang:2022icz} (see also Refs.~\cite{Burikham:2023bil, Oancea:2023ylb,Harko:2024fnt,Visa:2024fii,Molla:2024zxi} for related studies on Weyl gravity and their astrophysical and cosmological implications), and Solar System tests were examined in Ref.~\cite{Khodadi:2025gtq}, we aim at exploring these tests in light of the black hole solutions deduced in Ref.~\cite{Lima:2024cys}. Our goal is to derive observational constraints on the free parameter $\omega$ and assess the viability of the non-minimally coupled Weyl connection model in the weak-field regime. We further extend our analysis beyond the Solar System by testing the solutions against current observations of the S2 star near Sgr A*, thereby probing the strong-field regime and providing complementary constraints on the theory.

This work is organized as follows. In Sec. \ref{sec:BHsolutions} we overview the Schwarzschild-like black hole solutions in the model. We compute the gravitational redshift test in Sec. \ref{sec:gravitationalredshift}, the perihelion test in Sec. \ref{sec:perihelion}, the light deflection test in Sec. \ref{sec:lightdeflection}, and the radar echo delay in Sec. \ref{sec:radarecho}, assessing their implications for the model deriving constraints on the parameters. In Sec.~\ref{Sgr}, we test the black hole solutions using current observations of the S2 star, extending our analysis beyond the Solar System.
Finally, we draw our discussion and conclusions in Sec. \ref{sec:conclusion}.

\section{Vacuum Schwarzschild-like Solutions}\label{sec:BHsolutions}
In this section, we summarize and discuss the vacuum black hole solutions obtained in \cite{Lima:2024cys}. The study explores static spherically symmetric solutions in a modified gravity framework with a non-minimal coupling between matter and curvature, extended to incorporate a Weyl connection. 

This geometric setup introduces a vector field responsible for the non-metricity property, meaning the covariant derivative of the metric does not vanish but is proportional to the metric itself
\begin{equation}\label{D}
\bar{\nabla}_\lambda g_{\mu\nu} = A_\lambda g_{\mu\nu},
\end{equation}
where $A_\lambda$ is the Weyl vector field, $g_{\mu\nu}$ is the metric tensor, and the generalized covariant derivative is given by $\bar{\nabla}_\lambda g_{\mu\nu}=\nabla_\lambda g_{\mu\nu}-\bar{\bar{\Gamma}}^\rho_{\mu\lambda}g_{\rho\nu}-\bar{\bar{\Gamma}}^\rho_{\nu\lambda}g_{\rho\mu}$. Here, $\nabla_\lambda$ denotes the covariant derivative with Levi--Civita connection and $\bar{\bar{\Gamma}}^\rho_{\mu\nu}$ represents the additional connection terms encoding the Weyl non-metricity.

The Riemann tensor can be extended to incorporate the Weyl connection, considering ${\bar{\Gamma}^\rho_{\mu\nu}=\Gamma^\rho_{\mu\nu}+\bar{\bar{\Gamma}}^\rho_{\mu\nu}}$, such that
	\begin{equation}	\bar{R}^\rho_{\mu\sigma\nu}=\partial_\sigma \bar{\Gamma}^\rho_{\nu\mu}-\partial_\nu\bar{\Gamma}^\rho_{\sigma\mu}+\bar{\Gamma}^\rho_{\sigma\lambda}
		\bar{\Gamma}^\lambda_{\nu\mu}-\bar{\Gamma}^\rho_{\nu\lambda}\bar{\Gamma}^\lambda_{\sigma\mu}. \label{Riemann-total}
	\end{equation}
	
The generalized Ricci tensor is obtained by contracting the first and third indices of the curvature tensor, yielding
	\begin{equation}
		\begin{aligned}
		\bar{R}_{\mu\nu}=&R_{\mu\nu}+\frac{1}{2}A_\mu A_\nu +\frac{1}{2}g_{\mu\nu} \left(\nabla_\lambda-A_\lambda\right)A^\lambda +\tilde{F}_{\mu\nu}\\
		&+\frac{1}{2}\left(\nabla_\mu A_\nu+\nabla_\nu A_\mu\right)=R_{\mu\nu}+\bar{\bar{R}}_{\mu\nu},
		\label{Riccitensor}
	\end{aligned}
	\end{equation}
where $R_{\mu\nu}$ is the Ricci tensor constructed with the Levi--Civita connection and $\tilde{F}_{\mu\nu}=\partial_\mu A_\nu-\partial_\nu A_\mu$ is the strength tensor of the Weyl field. The trace of the generalized Ricci tensor, corresponding to the scalar curvature in the presence of the Weyl connection, is given by
	\begin{equation}
		\bar{R}=R+3\nabla_\lambda A^\lambda-\frac{3}{2}A_\lambda A^\lambda=R+\bar{\bar{R}},
	\end{equation}
where $R$ is the usual Ricci scalar. It is worth emphasizing that the generalized scalar curvature receives contributions from the metric-compatible part of the connection as well as from the Weyl non-metricity, which is encoded in the Weyl vector.

The action of the theory is given by \cite{Gomes:2018sbf}:
\begin{equation}
S = \int \left[ \kappa f_1(\bar{R}) + f_2(\bar{R}) \mathcal{L} \right] \sqrt{-g} \, d^4x,
\end{equation}
where $f_1(\bar{R})$ and $f_2(\bar{R})$ are function of the scalar curvature $\bar{R}$, built from the generalized connection.

Varying the action with respect to the vector field, up to boundary terms, it is possible to obtain the constraint-like equations\cite{Gomes:2018sbf}:
\begin{equation}
	\nabla_\lambda \Theta(\bar{R})=-A_\lambda \Theta(\bar{R}), 
	\label{eqconstraint}
\end{equation}
where $\Theta(\bar{R})=F_1(\bar{R})+F_2(\bar{R})\mathcal{L}$ and $F_i(\bar{R})=\frac{d f_i(\bar{R})}{d \bar{R}},$ $i\in \{1,2\}$. 

Varying the action with respect to the metric and taking into account the previous equation, we obtain the field equations\cite{Gomes:2018sbf}:
\begin{equation}
	\left( R_{\mu\nu}+\bar{\bar{R}}_{(\mu\nu)} \right) \Theta(\bar{R})-\frac{1}{2}g_{\mu\nu}f_1(\bar{R})=\frac{f_2(\bar{R})}{2}T_{\mu\nu},
	\label{fieldeqs}
\end{equation}
where parentheses denote symmetrization over the enclosed indices, and $T_{\mu\nu}$ is the energy-momentum tensor built from the matter Lagrangian, $T_{\mu\nu}=-\frac{2}{\sqrt{-g}}\frac{\delta (\sqrt{-g}\mathcal{L})}{\delta g^{\mu\nu}}$.

The trace of the metric field equations is
\begin{equation} 
	\Theta(\bar{R}) \bar{R}-2 f_1(\bar{R})=\frac{f_2(\bar{R})}{2}T,
	\label{trace}
\end{equation}
where $T=g^{\mu\nu}T_{\mu\nu}$.

In vacuum ($\mathcal{L}=0$), the model reduces to an $f(R)$-type theory with Weyl connection. Equation (\ref{trace}) impose that $f_1(\bar{R}) = \gamma \bar{R}^2$. Therefore, in \cite{Lima:2024cys}, vacuum solutions satisfying the following conditions were investigated:
\begin{subequations}
		\begin{align}
\bar{R}&=0,\\
\bar{R}_{(\mu\nu)}&=R_{\mu\nu} + \bar{\bar{R}}_{(\mu\nu)} = 0,
		\end{align}
\end{subequations}
where $\bar{R}_{(\mu\nu)}$ is the full generalized Ricci tensor, and $\bar{\bar{R}}_{(\mu\nu)}$ is the Weyl-induced correction, as given explicitly in Eq. (\ref{Riccitensor}).

By setting the following static line element in spherical coordinates
\begin{equation}\label{metric}
	ds^2 = -e^{\alpha(r)} dt^2 + e^{\beta(r)} dr^2 + r^2(d\theta^2 + \sin^2\theta \, d\phi^2),
\end{equation}
where \( g_{tt} = -e^{\alpha(r)} \) and \( g_{rr} = e^{\beta(r)} \), the Weyl vector is restricted by the symmetry of the field equations to only two possible ans\"atze: a purely radial configuration and a configuration with both temporal and radial components.

\subsection{Case 1: Purely Radial Weyl Vector}
The simplest ansatz is
\begin{equation}
A_\mu = (0, A(r), 0, 0), \quad A(r) \geq 0.
\end{equation}
Solving the field equations under the assumption $\beta(r) = -\alpha(r) + \epsilon$\footnote{Note on the constant \(\epsilon\): The quantity \(\epsilon\) appearing in the assumption \(\beta(r) = -\alpha(r) + \epsilon\) is a constant of integration that controls the relative normalization between the \(g_{tt}\) and \(g_{rr}\) components of the metric. It can be absorbed by a constant rescaling of the time coordinate, \(t \to e^{-\epsilon/2}t\), and therefore does not appear in the final solutions. Consequently, we set \(\epsilon = 0\) without loss of generality, recovering \(e^{\beta(r)} = e^{-\alpha(r)}\).} yields
\begin{equation}
A(r) = \frac{2}{r + \omega}.
\end{equation}
Mathematically, \(\omega\) arises as a constant of integration when solving the vacuum field equations for the assumed form of the Weyl vector. Physically, however, it plays the role of a free model parameter: it has dimensions of length, characterizes the strength of the Weyl connection, and its value is not predicted by the theory. Consequently, \(\omega\) must be constrained by observations, as we do throughout this work using Solar System tests and stellar orbits near Sgr~A\(^*\). This dual status is familiar from other scenarios---for instance, the cosmological constant in Schwarzschild--(anti-)de Sitter solutions or the charge in Reissner--Nordstr\"om black holes---where an integration constant of the vacuum solution simultaneously acts as a parameter of the theory. In the present context, the observational bounds on \(\omega\) therefore translate directly into constraints on the allowed strength of the Weyl connection.

The corresponding metric functions are
\begin{subequations}\label{sol1}
	\begin{align}
		e^{\alpha(r)} &= 1 - \frac{2M}{r} + \frac{2(\omega + 3M)}{\omega^2} r + \frac{\omega + 4M}{\omega^3} r^2, \\ 
		e^{\beta(r)} &= \frac{1 + 6M/\omega}{e^{\alpha(r)}}.	
	\end{align}
\end{subequations}

This solution admits an event horizon at ${r_H = 2M \frac{\omega}{4M + \omega}}$ and 
reduces to Schwarzschild near the black hole ($g_{tt} \simeq 1 - \frac{2M}{r}$). Asymptotically, however, it acquires linear and quadratic terms in $r$. The Ricci scalar is given by
\begin{equation}
R = -\frac{12(r+\omega)[(4M+\omega)r - M\omega]}{\omega^2(6M+\omega) r^2},
\end{equation}
and Kretschmann Invariant reads 
\begin{equation} 
K = \frac{48M^2}{r^6} \left( \frac{r+\omega}{6M+\omega} \right)^2.
\end{equation}
Both quantities diverge at $r=0$, indicating the presence of an essential curvature singularity at the origin, and remain finite at the horizon.

In the present work, we want to investigate the implications of the Schwarzchild-like black hole solutions in the Solar System regime and derive observational constraints on the parameter $\omega$. Since Solar System tests probe the weak-field limit of gravity, it is sufficient to consider the metric functions expanded to first order in the relevant small quantities. We therefore adopt the corresponding weak-field approximation of the above solution throughout our analysis.

Let us define the two dimensionless expansion parameters
\begin{equation}
\epsilon_1 = \frac{M}{r}  \quad \text{and} \quad \epsilon_2 = \frac{r}{\omega}.
\end{equation}
Throughout the Solar System, both quantities are expected to be small. For instance, in the Sun--Mercury system one has $\frac{M}{r} \sim 10^{-8}$, while the magnitude of $\frac{r}{\omega}$ will be constrained by observations. In what follows, we assume that these parameters are of comparable order,
\begin{equation}
\epsilon_1 \sim \epsilon_2 \sim \epsilon \ll 1,
\end{equation}
and retain only terms up to first order in $\epsilon$. Consequently:
\begin{eqnarray}
\frac{M}{\omega}& =  \epsilon_1 \epsilon_2 \sim \epsilon^2 \quad \text{(second order)}.\\
\frac{r^2}{\omega^2}& = {\epsilon_2}^2 \sim \epsilon^2 \quad \text{(second order)}.\\
\frac{Mr}{\omega^2}& =  \epsilon_1 {\epsilon_2}^2 \sim \epsilon^3 \quad \text{(third order)}.
\end{eqnarray}
Thus, to first order in $\epsilon$, we keep terms of order $\frac{M}{r}$ and $\frac{r}{\omega}$, and drop all terms of order $\epsilon^2$ or higher, including $\frac{M}{\omega}$, $\frac{r^2}{\omega^2}$, $\frac{Mr}{\omega^2}$, etc. This systematic expansion ensures consistency and avoids the spurious retention of second-order terms.

Retaining only the leading contributions, the solution (\ref{sol1}) reduces to
\begin{subequations}\label{eq:aprox_first}
	\begin{align}
		e^{\alpha(r)} &\simeq 1 - \frac{2M}{r} +\frac{2r}{\omega}, \\ 
		e^{\beta(r)} &\simeq \frac{1}{e^{\alpha(r)}}.	
	\end{align}
\end{subequations}

Since the quadratic term grows faster than the linear term as $r\longrightarrow \infty$, the linear term is subdominant and can be neglected in the asymptotic analysis, even though it dominates in the Solar System regime.

\subsection{Case 2: Time-Radial Weyl Vector}\label{B}
The second ansatz is
\begin{equation}
A_\mu = (A_0(r), A_1(r), 0, 0), \quad A_0(r) \neq 0,
\end{equation}
subject, by the symmetry of the field equations, to the constraint $A_0'(r) + (A_1(r) - \alpha'(r))A_0(r) = 0$. This constraint follows from the non-diagonal components of the field equations. In particular, imposing \(\bar{R}_{(\mu\nu)} = 0\) for \(\mu \neq \nu\) shows that the angular components of the Weyl vector must vanish and that, for the time-radial configuration, the temporal and radial components are related by the above constraint.

Solving the field equations yields
\begin{subequations}
	\begin{align}
		A_0(r) &= \frac{1}{\omega} \left(1 - \frac{2M}{r}\right), \\
		A_1(r) &= \frac{2r}{r^2 - 4\omega^2},
	\end{align}
\end{subequations}
with $\omega>0$.  The corresponding metric functions are given by
\begin{subequations}\label{sol2}
	\begin{align}
		e^{\alpha(r)} &= 1 - \frac{2M}{r} + \frac{M}{2\omega^2} r - \frac{1}{4\omega^2} r^2, \\
		e^{\beta(r)} &= \frac{1}{e^{\alpha(r)}}.
	\end{align}
\end{subequations}

This solution exhibits two horizons: $r_H^{(M)} = 2M$ and $r_H^{(\omega)} = 2\omega$, resembling black hole and cosmological horizons. The Ricci scalar is given by
\begin{equation}
R = \frac{3(r - M)}{\omega^2 r},\\ \label{R}
\end{equation}
and Kretschmann invariant takes the following form

\begin{equation}
	\begin{aligned}
		K = &\frac{48M}{r^6}\Bigg[
		\frac{M(47-2Mr+r^2)}
		{48} \\
		&-\frac{r^2\big[44M-2r(1+M^2)+Mr^2\big]}
		{192\omega^2}
		\Bigg].
	\end{aligned}
\end{equation}
Both quantities diverge at $r=0$, indicating the presence of an essential curvature singularity at the origin, and remain finite at the horizon.

As in the previous case, we focus on the weak-field regime relevant for Solar System tests. However, the structure of the present solution requires a different choice of expansion parameters. We therefore define
\begin{equation}
\epsilon_1 = \frac{M}{r} \quad \text{and} \quad \epsilon_2 = \frac{r^2}{\omega^2}.
\end{equation}

Unlike the previous solution, where the leading corrections to GR appeared at order $\frac{r}{\omega}$, the deviations in the present case arise only at order $\frac{r^2}{\omega^2}$. Consequently, this new $\epsilon_2$ provides the appropriate parameter controlling the Weyl-induced corrections.

We assume that these parameters are of comparable order, $\epsilon_1 \sim \epsilon_2 \sim \epsilon \ll 1$, and retain only terms up to first order in $\epsilon$. Consequently, mixed terms such as $\frac{M r}{\omega^2}=\epsilon_1 \epsilon_2 \sim \epsilon^2$ are of second order and will therefore be neglected. 

Retaining only the leading contributions, the solution (\ref{sol2}) reduces to
\begin{subequations}\label{eq:aprox_second}
	\begin{align}
		e^{\alpha(r)} &\simeq 1 - \frac{2M}{r} -\frac{r^2}{4\omega^2}, \\ 
		e^{\beta(r)} &= \frac{1}{e^{\alpha(r)}}.
	\end{align}
\end{subequations}
For Solution II, the linear term in $g_{tt}$ is $\frac{M}{2\omega^2} r$, which is suppressed by an additional factor of $\frac{M}{\omega}$ compared to the linear term in Solution I. Consequently, in the Solar System regime ($r \ll \omega$), the quadratic term $-\frac{r^2}{4\omega^2}$ dominates over the linear one, since
\begin{equation}
\frac{\text{Linear}}{\text{Quadratic}} = \frac{(M/2\omega^2)r}{r^2/(4\omega^2)} = \frac{2M}{r} \ll 1.
\end{equation}
This is the origin of the $\frac{1}{\omega^2}$ scaling of all Solar System observables for Solution II, in contrast to the $\frac{1}{\omega}$ scaling found for Solution I, where the linear term is not suppressed by $\frac{M}{\omega}$.

In both Solutions I and II \footnote{Henceforth, we shall refer to Cases 1 and 2 as Solutions I and II, respectively.}, the parameter $\omega$ has dimensions of length, and the Kretschmann invariant diverges only at $r=0$, confirming the presence of a central essential singularity.

\subsection{Asymptotic Behavior of Solutions I and II: Effective Cosmological Constant}

The asymptotic structure of the two Schwarzschild-like solutions reveals an intriguing connection with the Schwarzschild--(anti-)de Sitter family of spacetimes. In both cases, the metric contains a quadratic contribution proportional to $r^2$, a characteristic feature of black hole solutions in the presence of a cosmological constant. Although the origin of this term is fundamentally different, arising here from the Weyl connection rather than from an explicit cosmological constant in the gravitational action, its presence naturally motivates a comparison with the asymptotic behavior of Schwarzschild--(A)dS geometries. 

For Solution I, the quadratic term is given by $\frac{\omega+4M}{\omega^3} r^2$. Since the Solar System constraints obtained in this work require $\omega\gg M$, the mass-dependent contribution becomes negligible, yielding an asymptotic correction proportional to $\frac{r^2}{\omega^2}$. Interestingly, this hierarchy is also supported by an independent strong-field analysis of black hole shadows within the same theoretical framework, where compatibility with current Event Horizon Telescope observations was likewise found to favor the regime $\omega\gg M$ \cite{Gomes:2026qgm}. Consequently, both weak- and strong-field tests consistently indicate that the physically relevant sector of the theory satisfies $\omega$ much larger than the mass of the central object. Owing to its positive sign, the asymptotic behavior resembles that of a Schwarzschild--anti-de Sitter spacetime, although its physical origin is entirely different.

A similar interpretation applies to Solution II, whose the quadratic term,$-\frac{r^2}{4\omega^2}$, has the same sign as the Schwarzschild--de Sitter metric. Furthermore, this solution possesses a second horizon located at $r_H^{(\omega)}=2\omega$, in addition to the black hole event horizon at $r_H^{(M)}=2M$, closely resembling the coexistence of black hole and cosmological horizons in Schwarzschild--de Sitter spacetime.

Motivated by this asymptotic correspondence, one may formally define an effective cosmological constant by matching the quadratic term with the Schwarzschild--(anti-)de Sitter metric. Neglecting the contribution proportional to $\frac{M}{\omega}$, the resulting effective cosmological constants become 
\begin{equation}
	\Lambda_\text{eff}^{\text{(I)}}\simeq -\frac{3}{\omega^2} \quad \text{and} \quad \Lambda_\text{eff}^{\text{(II)}}=\frac{3}{4\omega^2}, 
\end{equation}
where the opposite signs reflect the asymptotic anti-de Sitter-like and de Sitter-like behaviors of Solutions I and II, respectively.

\section{Gravitational redshift}\label{sec:gravitationalredshift}

For a static, spherically symmetric metric (\ref{metric}), the gravitational redshift is determined by the time dilation between two fixed positions \(r_1\) and \(r_2\). For an observer at \(r_2\) receiving light emitted at \(r_1\), the frequency ratio is
\begin{equation}
	\frac{\nu_2}{\nu_1} = \sqrt{\frac{g_{tt}(r_1)}{g_{tt}(r_2)}} = \sqrt{\frac{e^{\alpha(r_1)}}{e^{\alpha(r_2)}}}.
\end{equation}
The redshift parameter \(z\) is such that
\begin{equation}
	1 + z = \frac{\nu_1}{\nu_2} = \sqrt{\frac{e^{\alpha(r_2)}}{e^{\alpha(r_1)}}}.
\end{equation}
\vspace{0.1cm}

For metric Solution I, considering the first-order approximation (\ref{eq:aprox_first}), with \(r_1\) (emitter) and \(r_2\) (receiver), we have
\begin{equation}
	1 + z \simeq \sqrt{\frac{1 - \frac{2M}{r_2} + \frac{2 r_2}{\omega}}
		{1 - \frac{2M}{r_1} + \frac{2r_1}{\omega}}}.
\end{equation}
Using the first-order approximations \(\sqrt{1+2x} \simeq 1 + x\) and \(\frac{1}{\sqrt{1+2x}} \simeq 1 - x\), for small \(x\), we get
\begin{equation}
	1 + z \simeq \left[ 1 - \frac{M}{r_2} + \frac{r_2}{\omega} \right] \left[ 1 + \frac{M}{r_1} - \frac{r_1}{\omega} \right],
\end{equation}
where we kept only the leading corrections. Expanding and retaining only terms up to first order in the previously defined small parameters, we obtain
\begin{equation}
	1 + z \simeq 1 + M\left(\frac{1}{r_1} - \frac{1}{r_2}\right) + \frac{1}{\omega}(r_2 - r_1).
	\label{eq:redshift_I_firstorder}
\end{equation}
The third term in Eq.~\eqref{eq:redshift_I_firstorder} is the leading correction due to the Weyl constant \(\omega\). The standard Schwarzschild redshift is therefore
\begin{equation}
	z_{\mathrm{GR}} = M\left(\frac{1}{r_1} - \frac{1}{r_2}\right),
	\label{eq:schwarzschild_redshift}
\end{equation}
and the extra redshift from the \(\omega\) term is
\begin{equation}
	\delta z_\omega = \frac{r_2 - r_1}{\omega}.
	\label{eq:extra_redshift_I}
\end{equation}
For the Solar System application, we take ${r_1 = R_\odot \simeq 6.96 \times 10^8 \, \mathrm{m}}$ as the photon emission point at the Sun's surface, and ${r_2 = 1 \, \mathrm{AU} \simeq 1.496 \times 10^{11} \, \mathrm{m}}$ as the reception point at Earth. As a result
\begin{equation}
	\delta z_\omega \simeq \frac{1.489 \times 10^{11}}{\omega}.
\end{equation}
The gravitational redshift predicted by GR for a photon emitted from the Sun's surface is \({z_{\text{GR}} \simeq 2.12 \times 10^{-6}}\). Observations (after correcting for Doppler and convective effects) agree with this prediction to within about \(1\%\) corresponding to an uncertainty of \(\delta z_{\text{max}} \simeq 2.12 \times 10^{-8}\) \cite{Shapiro:2004zz}.
Therefore, the correction to the GR prediction must satisfy 
\begin{equation}
	\frac{1.489 \times 10^{11}}{\omega} \lesssim 2.12 \times 10^{-8} \hspace{0.1cm}  \Rightarrow \hspace{0.1cm}  \omega \gtrsim 7.02 \times 10^{18}\ \text{m},
\end{equation}
which, in geometric units, considering ${M=M_\odot\simeq 1.477\times 10^3\text{m}}$ \cite{Will:2014kxa},  reads
\begin{equation} 
\frac{\omega}{M_\odot} \gtrsim 4.75 \times 10^{15}.\\
\end{equation}

\vspace{0.25cm}

For metric Solution II, considering the first-order approximation (\ref{eq:aprox_second}), the redshift is given by
\begin{equation}
	1 + z \simeq \sqrt{\frac{1 - \frac{2M}{r_2}  - \frac{1}{4\omega^2} {r_2}^2}
		{1 - \frac{2M}{r_1}  - \frac{1}{4\omega^2} {r_1}^2}}.
\end{equation}
This leads to
\begin{equation}
	1 + z \simeq \left[ 1 - \frac{M}{r_2}  - \frac{1}{8\omega^2} {r_2}^2 \right] 
	\left[ 1 + \frac{M}{r_1}+ \frac{1}{8\omega^2} {r_1}^2 \right],
\end{equation}
where we kept only the leading corrections. Expanding and retaining only terms up to first order in the previously defined small parameters, we obtain
\begin{equation}
	1 + z \simeq 1 + M\left(\frac{1}{r_1} - \frac{1}{r_2}\right) 
	- \frac{1}{8\omega^2}({r_2}^2 - {r_1}^2).
	\label{eq:redshift_II_firstorder}
\end{equation}
Hence, the additional contribution to the redshift due to \(\omega\) is
\begin{equation}\label{z}
	\delta z_\omega = -\frac{1}{8\omega^2}({r_2}^2 - {r_1}^2).
\end{equation}
Applying the same definitions $r_1 = R_\odot$ and ${r_2 = 1\,\mathrm{AU}}$, yields
\begin{equation}
	\delta z_\omega \simeq -\frac{2.80 \times 10^{21}}{\omega^2}.
\end{equation}
Requiring this correction to remain below the observational uncertainty \(\delta z_{\text{max}} \simeq 2.12 \times 10^{-8}\), we obtain
\begin{equation}
	\frac{2.80 \times 10^{21}}{\omega^2} \lesssim 2.12 \times 10^{-8} \hspace{0.1cm} \Rightarrow \hspace{0.1cm}  \omega \gtrsim 3.63 \times 10^{14}\ \text{m},
\end{equation}
which, in geometric units, reads 
\begin{equation} 
\frac{\omega}{M_\odot} \gtrsim 2.53 \times 10^{11}. 
\end{equation}

\vspace{0.25cm}

The two solutions lead to qualitatively different corrections to the gravitational redshift. For Solution I, the Weyl contribution is positive and therefore increases the gravitational redshift relative to the standard GR prediction. In contrast, the correction associated with Solution II is negative, leading to a reduction of the predicted redshift. Although the correction in Solution II scales as \(\frac{1}{\omega^2}\), the large factor \((r_2^2-r_1^2)\) significantly enhances its magnitude. As a result, the gravitational redshift test still provides a meaningful constraint, \({\frac{\omega}{M_\odot}\gtrsim2.53\times10^{11}}\). Nevertheless, this bound remains substantially weaker than that obtained for Solution I, \({\frac{\omega}{M_\odot}\gtrsim4.75\times10^{15}}\), differing by approximately four orders of magnitude.

\section{Perihelion shift}\label{sec:perihelion}

Before proceeding with the geodesic analysis, we clarify the treatment of the normalization condition in the presence of non-metricity. In Weyl geometry, the connection satisfies $\bar{\nabla}_\lambda g_{\mu\nu} = A_\lambda g_{\mu\nu}$ [Eq.~\eqref{D}], implying that lengths are not conserved under parallel transport. Consequently, for autoparallel curves of the Weyl connection, the magnitude of the four-velocity is not constant; instead one has $g_{\mu\nu}\frac{\mathrm{d} x^\mu}{\mathrm{d}\lambda}\frac{\mathrm{d} x^\nu}{\mathrm{d}\lambda} = -l^2(\lambda)$, where $\lambda$ is the affine parameter, related to the proper time $\tau$ via $\mathrm{d}\tau = l(\lambda)\, \mathrm{d}\lambda$ (see, e.g., Ref.~\cite{Iosifidis:2018diy}).  

In this work, we assume that test particles follow geodesics of the Levi-Civita connection, while the Weyl vector enters only at the level of the gravitational field equations through the scalar curvature $\bar{R}$. This approach is consistent with the Palatini-like formulation in which the matter action is constructed from the metric alone and does not couple directly to the Weyl connection. Under this assumption, the standard normalization $g_{\mu\nu}\frac{\mathrm{d} x^\mu}{\mathrm{d}\tau}\frac{\mathrm{d} x^\nu}{\mathrm{d}\tau}~=~-1$ holds, and the affine parameter may be identified with proper time.\\

In the equatorial plane ($\theta = \pi/2$), the geodesic Lagrangian is
\begin{equation}
	\mathcal{L} = \frac12 \left[ -e^{\alpha(r)} \left( \frac{\mathrm{d} t}{\mathrm{d} \tau} \right)^2 + e^{\beta(r)} \left( \frac{\mathrm{d} r}{\mathrm{d} \tau} \right)^2 + r^2 \left( \frac{\mathrm{d} \phi}{\mathrm{d} \tau} \right)^2 \right].
	\label{eq:lagrangian}
\end{equation}
The Euler--Lagrange equations yield conserved quantities
\begin{subequations}\label{eq:qts_conserved}
	\begin{align}
		E &= \frac{e^{\alpha(r)} }{f_2(0)} \frac{\mathrm{d} t}{\mathrm{d} \tau} = \text{constant}, \label{eq:E_conserved}\\
		L &= \frac{r^2}{f_2(0)}\frac{\mathrm{d} \phi}{\mathrm{d} \tau} = \text{constant}, \label{eq:L_conserved}
	\end{align}
\end{subequations}
where without loss of generality we shall set ${f_2(0)=1}$ as in the weak-field regime with fifth-force effects and non-minimal couplings, via Yukawa terms, are extremely constrained \cite{Adelberger:2003zx,Wolf:2025jed,Geng:2015nnb,March:2021mqu}.

The normalization condition for timelike geodesics (test particles following Levi-Civita geodesics) is $g_{\mu\nu}\frac{\mathrm{d} x^\mu}{\mathrm{d} \tau}\frac{\mathrm{d} x^\nu}{\mathrm{d} \tau}~=~-1$, which expands to
\begin{equation}
	-e^{\alpha(r)}\left( \frac{\mathrm{d} t}{\mathrm{d} \tau} \right)^2 + e^{\beta(r)}\left( \frac{\mathrm{d} r}{\mathrm{d} \tau} \right)^2 + r^2\left( \frac{\mathrm{d} \phi}{\mathrm{d} \tau} \right)^2 = -1.
	\label{eq:normalization}
\end{equation}

Using the relations (\ref{eq:qts_conserved}) and noting that ${e^{\beta(r)}=e^{-\alpha(r)}}$ (which holds approximately for Solution I), the previous expression can be rewritten as 
\begin{equation}
	\left( \frac{\mathrm{d} r}{\mathrm{d} \tau} \right)^2 = E^2 - e^{\alpha}\left(1 + \frac{L^2}{r^2}\right).
	\label{eq:radial_final}
\end{equation}
Defining $u = \frac{1}{r}$, it follows that
\begin{equation}
\frac{\mathrm{d} r}{\mathrm{d}\tau} = \frac{\mathrm{d} r}{\mathrm{d}\phi}\frac{\mathrm{d}\phi}{\mathrm{d}\tau} = \frac{\mathrm{d} r}{\mathrm{d} \phi}\frac{L}{r^2} = -L\frac{\mathrm{d} u}{\mathrm{d}\phi},\label{eq:du}
\end{equation}
where we used $\frac{\mathrm{d} r}{\mathrm{d} \phi}=-\frac{1}{u^2} \frac{\mathrm{d} u}{\mathrm{d} \phi}$. Substituting this result into Eq.~\eqref{eq:radial_final}, we obtain
\begin{equation}
	L^2\left(\frac{\mathrm{d}u}{\mathrm{d}\phi}\right)^2 = E^2 - e^{\alpha}\left(1 + L^2 u^2\right).
	\label{eq:orbit_u}
\end{equation}
At the closest approach $r = r_0$ (perihelion), we have $\frac{\mathrm{d} u}{\mathrm{d} \phi}=0$ and $u = u_0 = \frac{1}{r_0}$, then yields
\begin{equation}
	E^2 = e^{\alpha(u_0)}\left(1 + L^2 {u_0}^2\right).
	\label{eq:E_turning}
\end{equation}
Replacing Eq.~\eqref{eq:E_turning} back into Eq.~\eqref{eq:orbit_u}, we find
\begin{equation}
	L^2\left(\frac{\mathrm{d}u}{\mathrm{d}\phi}\right)^2 = e^{\alpha(u_0)}\left(1 + L^2 {u_0}^2\right) - e^{\alpha(u)}\left(1 + L^2 u^2\right). 
	\label{eq:orbit_turning}
\end{equation}

Considering Solution I, and using the relation ${u=\frac{1}{r}}$, we have $e^{\alpha(u)}\simeq 1-2M u+\frac{2}{\omega u}$. Thus, Eq. (\ref{eq:orbit_turning}) can be rewritten as
\begin{equation}
\begin{aligned}
\hspace{-1cm}	\left(\frac{\mathrm{d} u}{\mathrm{d}\phi}\right)^2 &= ({u_0}^2 - u^2) - \frac{2M}{L^2}(u_0 - u) - 2M({u_0}^3 - u^3) +\\
	&+ \frac{2}{\omega L^2}\left(\frac{1}{u_0} - \frac{1}{u}\right) + \frac{2}{\omega}(u_0 - u).
	\label{eq:orbit_final}
\end{aligned}
\end{equation}
Differentiating both sides with respect to \(\phi\) and assuming \(\frac{du}{d\phi} \neq 0\), we arrive at the orbit equation
\begin{equation}
	\frac{\mathrm{d}^2 u}{\mathrm{d} \phi^2} + u = \frac{M}{L^2} + 3M u^2 + \frac{1}{\omega L^2 u^2} - \frac{1}{\omega}.
	\label{eq:orbit_correct}
\end{equation}
The terms $\frac{1}{\omega L^2 u^2} = \frac{r^2}{\omega L^2}$ and $-\frac{1}{\omega}$ are the leading Weyl corrections. In the limit $\omega \to \infty$, we recover the standard Schwarzschild equation
\begin{equation}
\frac{\mathrm{d}^2 u}{\mathrm{d}\phi^2} + u = \frac{M}{L^2} + 3M u^2.
\end{equation}

For a nearly circular orbit, we write $u = u_0(1 + \xi)$, with $\xi \ll 1$ and $u_0 = \frac{1}{p}$. Here, $p = a(1-e^2)$ is the semi-latus rectum, $a$ is the semi-major axis, and $e$ is the orbital eccentricity. In the Newtonian limit, the angular momentum is related to $p$ and $M$ through
\begin{equation}
	L^2 = M p.
	\label{eq:L2_Newton}
\end{equation}
We also note that for a circular orbit, the Newtonian equation $\frac{d^2u}{d\phi^2} + u =\frac{M}{L^2}$ gives $u_0 = \frac{M}{L^2} = \frac{1}{p}$, consistent with Eq.~\eqref{eq:L2_Newton}.

With these assumptions, substituting \({u=u_0(1+\xi})\) into Eq.~\eqref{eq:orbit_correct}, each term can be expanded to first order in \(\xi\) as follows:
\begin{equation}
	u \simeq u_0(1+\xi),~~~~~	u^2 \simeq {u_0}^2(1+2\xi),~~~	\frac{1}{u^2} \simeq \frac{1}{{u_0}^2}(1 - 2\xi).
\end{equation}
Using these expressions in Eq.~\eqref{eq:orbit_correct}, we obtain
\begin{equation}
\frac{\mathrm{d}^2 \xi}{\mathrm{d}\phi^2} + \left(1 - \frac{6M}{p} + \frac{2p^2}{\omega M}\right)\xi =3 \frac{M}{p}+\frac{p^2}{\omega M} -\frac{p}{\omega} .
\end{equation}
Defining $k^2 = 1 - \frac{6M}{p} + \frac{2p^2}{\omega M}$, the equation takes the form
\begin{equation} 
\frac{\mathrm{d}^2\xi}{\mathrm{d}\phi^2} + k^2 \xi=\text{constant}.  
\end{equation}
From the solution of the above equation, the precession of Mercury perihelion per orbit is therefore 
\begin{equation}\label{p}
\Delta\phi = 2\pi\left(\frac{1}{k} - 1\right) \simeq 2\pi\left(\frac{3M}{p} - \frac{p^2}{\omega M}\right),
\end{equation}
where the approximation follows from expanding $\frac{1}{k}$ to first order in $\frac{M}{p}$ and $\frac{p^2}{\omega M}$.

The first term reproduces the standard GR result, $\Delta\phi_{\text{GR}} = 6\pi \frac{M}{p}$. The additional contribution due to the Weyl term is therefore
\begin{equation}\label{PerihelionCorrectionI}
\delta(\Delta\phi)_{\omega} = -\frac{2\pi p^2}{\omega M} \quad \text{radians per orbit}.
\end{equation}
The negative sign indicates that the Weyl contribution opposes the standard Schwarzschild advance of the perihelion, thereby reducing the total relativistic precession. In this sense, the Weyl correction contributes a retrograde component to the orbital motion.  

We now derive a numerical constraint from Mercury's perihelion precession. For Mercury, we adopt the following parameters (see, e.g., Ref.~\cite{Will:2014kxa}): the solar mass in geometric units, $M_\odot = GM_\odot/c^2 \simeq 1.4766 \times 10^3$ m;  the semi-major axis, $a \simeq 5.7909 \times 10^{10}$ m; the orbital eccentricity, $e \simeq 0.20563$; the semi-latus rectum, $p = a(1-e^2) \simeq 5.55 \times 10^{10}$ m. The orbital period is $T \simeq 87.969$ days, corresponding to  $N \simeq 36525/87.969 \simeq 415.2$ orbits per century. The GR precession is $42.98''$ per century, while the observational accuracy (from radio tracking of Mercury, e.g., MESSENGER mission) is $\lesssim 0.001''$ per century~\cite{Park:2017zgd}.  Finally, we use the conversion factor $1 \text{ radian} = 206264.806''$.

In this case, the additional precession per century, in radians, is given by
\begin{equation}
|\delta(\Delta\phi)_{\text{cent, rad}}| = N \cdot \frac{2\pi p^2}{\omega M_\odot}= \frac{5.441 \times 10^{21}}{\omega}.
\end{equation}
Now, convert to arcseconds per century, we obtain 
\begin{equation}
|\delta(\Delta\phi)_{\text{cent, arcsec}}| = \frac{1.122 \times 10^{27}}{\omega}.
\end{equation}
Requiring this to be less than the observational accuracy of $0.001''$ per century, it follows 
\begin{equation}
\frac{1.122 \times 10^{27}}{\omega} \lesssim 10^{-3} \hspace{0.1cm} \Rightarrow\hspace{0.1cm}  \omega \gtrsim 1.12 \times 10^{30} \ \text{m},
\end{equation}
which, in geometric units, reads 
\begin{equation}
\frac{\omega}{M_\odot} \gtrsim 7.60 \times 10^{26}.
\end{equation}
\vspace{0.25cm}

Lets us now analyze the Solution II. Following the same procedure as before and using the approximation ${e^{\alpha(r)}\simeq1-2Mu-\frac{1}{4\omega^2 u^2}}$, Eq.~(\ref{eq:orbit_turning}) can be rewritten as 
\begin{equation}
	\begin{aligned}
	\left(\frac{\mathrm{d} u}{\mathrm{d}\phi}\right)^2&=({u_0}^2-u^2)-2M\left( {u_0}^3-u^3 \right)-\\
&-\frac{2M}{L^2}(u_0-u)-\frac{1}{4\omega^2L^2}\left(\frac{1}{{u_0}^2}-\frac{1}{u^2}\right). 
	\end{aligned}
\end{equation}
Differentiating both sides with respect to \(\phi\) and assuming \(\frac{du}{d\phi} \neq 0\), we arrive at the orbit equation
\begin{equation}
	\frac{\mathrm{d}^2 u}{\mathrm{d}\phi^2} + u = \frac{M}{L^2} + 3M u^2  - \frac{1}{4\omega^2 L^2 u^3}.
	\label{eq:orbit_II_original}
\end{equation}
The term $- \frac{1}{4\omega^2 L^2 u^3} = - \frac{r^3}{4\omega^2 L^2}$ is the leading Weyl correction. As expected, in the limit $\omega \to \infty$, we recover again the standard Schwarzschild equation. 

Applying the same procedure as before and expanding about the Newtonian circular orbit $u_0 = \frac{1}{p}$, we find that the squared angular frequency is
\begin{equation}
	k^2 =1-\frac{6M}{p} -\frac{3p^3}{4M\omega^2}.
	\label{eq:dk2_II}
\end{equation}
Therefore, the perihelion precession per orbit is given by
\begin{equation}
	\Delta \phi \simeq 2\pi \left( \frac{3M}{p}+\frac{3 p^3}{8 \omega^2 M} \right). 
	\label{eq:delta_precess_II}
\end{equation}
As before, the first term reproduces the standard GR result and the additional contribution due the Weyl term is such that 
\begin{equation}
	\delta (\Delta \phi)_\omega= \frac{3\pi p^3}{4\omega^2 M} \quad \text{radians per orbit.}
	\label{eq:delta_precess_II_dom}
\end{equation}
Since the Weyl correction enters with a positive sign, it reinforces the standard Schwarzschild precession, corresponding to an additional prograde contribution to the perihelion motion.

Using the same Mercury parameters introduced above, the additional perihelion precession per century, in radians, is given by
\begin{equation}
	\delta( \Delta \phi)_\text{cent, rad} =N.\frac{3\pi p^3}{4 \omega^3 M_\odot} = \frac{1.132 \times 10^{32}}{\omega^2}.
\end{equation}
Now, convert to arcseconds per century, we obtain 
\begin{equation}
	\delta( \Delta \phi)_\text{cent, arcsec}=\frac{2.335\times 10^{37}}{\omega^2}.
\end{equation}
Requiring this less than the observational accuracy of $0.001''$ per century, it follows
\begin{equation}
	\frac{2.334\times 10^{37}}{\omega^2}  \lesssim 10^{-3}   \hspace{0.1cm} \Rightarrow\hspace{0.1cm}  \omega \gtrsim  1.53\times 10^{20} \text{m},
\end{equation}
which, in geometric units, reads 
\begin{equation}
	\frac{\omega}{M_\odot} \gtrsim 1.04 \times 10^{17}. 
\end{equation}
The two solutions also lead to qualitatively different corrections to the perihelion precession. For Solution I, the Weyl correction has a negative sign and therefore opposes the standard Schwarzschild advance of the perihelion, reducing the total relativistic precession. In contrast, the correction associated with Solution II is positive and contributes an additional prograde component to the perihelion motion, enhancing the GR prediction.

Although the correction in Solution II scales as \(\frac{1}{\omega^2}\), the large factor \(\frac{p^3}{M}\) significantly amplifies its contribution. As a result, Mercury's perihelion precession still yields a nontrivial constraint, \(\frac{\omega}{M_\odot} \gtrsim 1.04\times10^{17}\). Nevertheless, this bound remains substantially weaker than that obtained for Solution I, \(\frac{\omega}{M_\odot} \gtrsim 7.60\times10^{26}\), differing by approximately ten orders of magnitude.

\subsection{Choice of Parameterization}

The results obtained above were derived by parametrizing the particle trajectories in terms of the proper time. In a metric theory, this choice is natural and leads directly to the usual geodesic equation. However, as mentioned before, in the presence of non-metricity the proper time and the affine parameter associated with autoparallel transport cannot be equivalent. It is therefore important to examine whether the choice of parameter affects the resulting orbital dynamics and, consequently, the bounds obtained from the perihelion shift.

To this end, we consider the motion of test particles using an affine parameter associated with the autoparallel transport defined by the Weyl connection. We impose the normalization condition ${g_{\mu\nu} \frac{\mathrm{d} x^\mu}{\mathrm{d} \lambda} \frac{\mathrm{d} x^\nu}{\mathrm{d}\lambda}=-\ell^2}$, where $\ell$ is, in general, a non-constant function along the trajectory. For autoparallel transport, the evolution of the norm of the tangent vector is governed by the Weyl vector, yielding the direct relation 
\begin{equation} 
\frac{\mathrm{d} (\ell^2)}{\mathrm{d} \lambda}=(A_\mu u^\mu) (\ell^2), \label{relation-l2}
\end{equation}
where $u^\mu=\frac{\mathrm{d} x^\mu}{\mathrm{d}\lambda}$ denotes the tangent vector to the particle's trajectory.

For Solution I, integrating the corresponding Weyl vector contribution gives an expression for $\ell^2$ containing an integration constant. We fix this constant by requiring $\ell^2\to1$ in the limit $r\to0$, which gives 
\begin{equation}
	\ell^2=\left(1+\frac{r}{\omega}\right)^2. 
\end{equation}
The resulting expression introduces an additional contribution to the orbital dynamics that was not present in the proper time parametrization considered above. Since the analysis is performed in the regime $\frac{r}{\omega}\ll 1$, this function can be expanded to the first order as 
\begin{equation}
	\ell^2 \simeq 1+\frac{2 r}{\omega}.
\end{equation}
Thus, the choice of an affine parameter introduces an additional contribution. We can now repeat the analysis performed above, imposing the normalization condition $u^\mu u_\mu=-\ell^2$. With this modification, Eq.~(\ref{eq:radial_final}) takes the form
\begin{equation} 
	\left( \frac{\mathrm{d} r}{\mathrm{d} \lambda} \right)^2=E^2-e^{\alpha(r)}\left( \ell^2+\frac{L^2}{r^2} \right), 
\end{equation}
where the conserved quantities are defined consistently with the chosen parametrization. Following the same procedure described above, the additional contribution arising from the Weyl term is
\begin{equation}
	\delta (\Delta \phi)_\omega = -\frac{4 \pi p^2}{\omega M} \text{radians per orbit}. 
\end{equation}
Comparing this result with Eq.~(\ref{PerihelionCorrectionI}), we find that the two expressions differ only by a factor of two. Therefore, the use of proper time in the preceding analysis does not appear to affect the qualitative conclusions of this work. The corresponding affine-parameter treatment leads to the same dependence on the Weyl parameter and changes the predicted corrections only by a numerical factor of order unity. Consequently, the observational bounds obtained above remain robust against this ambiguity in the choice of parametrization.\\ 
\vspace{0.25cm}

For Solution II, the expression for $\ell^2$  cannot be obtained directly in closed form. In this case, the Weyl vector is not locally exact, satisfying $\partial_r A_0\neq \partial_t A_1$,  so that the integral determining $\ell^2$ depends explicitly on the path followed by the particle. 

We therefore consider the angular component of the autoparallel transport equation, $u^\nu \bar{\nabla}_\nu u^\mu=0$, which is given by 
\begin{equation}
	\frac{\mathrm{d}^2 \phi}{\mathrm{d} \lambda^2 }+\left( \frac{2M-r}{\omega r}\frac{\mathrm{d} t}{\mathrm{d} \lambda}+\frac{8 \omega^2}{r(4\omega^2-r^2)}\frac{\mathrm{d}r}{\mathrm{d}\lambda} \right) \frac{\mathrm{d} \phi}{\mathrm{d} \lambda}. 
\end{equation}
Please note that $\bar{\nabla}$ denotes the covariant derivative associated with the generalized connection, and that the tangent vector in this analysis is given by 
${u^\mu=\left( \frac{\mathrm{d} t}{\mathrm{d}\lambda}, \frac{\mathrm{d} r}{\mathrm{d}\lambda},0,\frac{\mathrm{d} \phi}{\mathrm{d} \lambda} \right)}$, in the equatorial plane.

Using the conserved quantity, which gives ${\frac{\mathrm{d}\phi}{\mathrm{d}\lambda}=\frac{L}{r^2}}$, we obtain the relation
\begin{equation}
	\frac{\mathrm{d}t}{\mathrm{d}\lambda}=-\frac{2\omega r^2}{(4\omega^2-r^2)(2M-r)}\frac{\mathrm{d}r}{\mathrm{d}\lambda}. 
\end{equation}
Using this relation, it can be shown that
\begin{equation}
	A_\mu u^\mu= A_0(r) \frac{\mathrm{d} t}{\mathrm{d}\lambda}+A_1(r)\frac{\mathrm{d} r}{\mathrm{d}\lambda}=0. 
\end{equation}

Thus, from Eq.~(\ref{relation-l2}), we find that $\frac{\mathrm{d} (\ell^2)}{\mathrm{d}\lambda}=0$, implying that $\ell^2$ is constant. We can therefore normalize it to unity, recovering $u^\mu u_\mu=-1$. Hence, for Solution II, the proper time and affine-parameter descriptions can be identified in the weak-field regime.

\section{Light Deflection from the Orbit Equation}
\label{sec:lightdeflection}

We now apply the consistent first-order expansion to null geodesics. For null geodesics, the normalization condition is \(\mathrm{d}s^2=0\). Following the same steps as in the perihelion precession calculation, we obtain the modifications to the light-deflection angle.

Similarly, we have
\begin{equation}
	-e^{\alpha(r)}\left( \frac{\mathrm{d} t}{\mathrm{d} \lambda} \right)^2+e^{\beta(r)}\left( \frac{\mathrm{d} r}{\mathrm{d} \lambda} \right)^2+r^2\left( \frac{\mathrm{d} \phi}{\mathrm{d} \lambda} \right)^2=0.
\end{equation}

Using the relations (\ref{eq:qts_conserved}) and noting that ${e^{\beta(r)}=e^{-\alpha(r)}}$ (which holds approximately for Solution I), the previous expression can be rewritten as 
\begin{equation}
	\left( \frac{\mathrm{d} r}{\mathrm{d} \lambda} \right)^2=E^2-e^{\alpha(r)}\frac{L^2}{r^2}. 
\end{equation}

Introducing once again the variable $u=\frac{1}{r}$ and use the relation (\ref{eq:du}), we proceed as in the timelike-geodesic case. At the point of closest approach, \(r=r_0\), we have \(\frac{\mathrm{d} u}{\mathrm{d}\phi}=0\), which implies ${E^2=e^{\alpha(u_0)}L^2 {u_0}^2}$. Using this result, we obtain
\begin{equation}
	\left(\frac{\mathrm{d} u}{\mathrm{d}\phi}\right)^2 = e^{\alpha(u_0)}{u_0}^2-e^{\alpha(u)}u^2. 
	\label{eq:null_simplified}
\end{equation}
\vspace{0.1cm}

Considering Solution I, Eq. (\ref{eq:null_simplified}) takes the form
\begin{equation}
	\left(\frac{\mathrm{d}u}{\mathrm{d}\phi}\right)^2 = ({u_0}^2 - u^2) - 2M({u_0}^3 - u^3) + \frac{2}{\omega}(u_0 - u).	\label{eq:null_expanded}
\end{equation}
Differentiating both sides with respect do $\phi$ and assuming $\frac{du}{d\phi}\neq0$, we arrive at the null orbit equation
\begin{equation}
	\frac{\mathrm{d}^2 u}{\mathrm{d}\phi^2} + u = 3M u^2 - \frac{1}{\omega}.
	\label{eq:null_orbit_correct}
\end{equation}
In the limit $\omega \to \infty$, we recover the standard Schwarzschild null equation $\frac{d^2 u}{d \phi^2} + u = 3M u^2$.

Considering $u = u^{(0)} + u^{(1)}$, where ${u^{(0)} = u_0 \cos\phi}$ is the unperturbed straight-line solution (with $u_0=\frac{1}{r_0}$ the inverse impact parameter), substituting into Eq.~\eqref{eq:null_orbit_correct} and retaining only terms linear in the perturbation \(u^{(1)}\), while using the relation $\cos^2\phi = \frac12(1 + \cos 2\phi)$, we obtain
\begin{equation}
	\frac{\mathrm{d}^2 u^{(1)}}{\mathrm{d}\phi^2} + u^{(1)} =  \frac{3}{2}M {u_0}^2 + \frac{3}{2}M {u_0}^2 \cos 2\phi - \frac{1}{\omega}.
	\label{eq:null_perturb}
\end{equation}
We seek a particular solution of the form $u^{(1)} = A + B\cos 2\phi$. Matching the constant and oscillatory terms yields $A=\frac{3Mu_0^2}{2}-\frac{1}{\omega}$ and $B=-\frac{Mu_0^2}{2}$. Therefore,
\begin{equation}
	u^{(1)}(\phi) = \frac{3}{2}M {u_0}^2 - \frac{1}{\omega} - \frac{1}{2}M {u_0}^2 \cos 2\phi.
	\label{eq:null_u1}
\end{equation}
The full solution is given by $u(\phi) = u_0\cos\phi + u^{(1)}(\phi)$. 

The unperturbed trajectory crosses $u=0$ at $\phi=\pm\frac{\pi}{2}$. Due to perturbations, the crossing occurs at ${\phi = \pm\left(\frac{\pi}{2} + \epsilon\right)}$, with $\epsilon$ small. Without loss of generality, let us set $\phi = \frac{\pi}{2} + \epsilon$. Then, considering  the first-order approximation $\cos\left( \pi +2\epsilon \right)\simeq -1$ and solving the equation $u(\phi)=0$, we can obtain
\begin{equation}
	\epsilon = 2M u_0 - \frac{1}{\omega u_0}.
	\label{eq:null_epsilon}
\end{equation}
The total deflection angle (beyond $\pi$) is $\Delta\phi = 2\epsilon$, therefore
\begin{equation}
	\Delta\phi = 4M u_0 - \frac{2}{\omega u_0}.
	\label{eq:null_deflection}
\end{equation}
In terms of the impact parameter $r_0 = 1/u_0$ we have
\begin{equation}
	\Delta\phi = \frac{4M}{r_0} - \frac{2r_0}{\omega}.
	\label{eq:deflection_correct}
\end{equation}
The standard GR deflection is $\Delta\phi_{\text{GR}}=\frac{4M}{r_0}$ and the Weyl first-order correction is 
\begin{equation}
\delta (\Delta\phi)_\omega=-\frac{2r_0}{\omega}.
\end{equation}

The deflection of a starlight by the Sun has been measured to high precision using very-long-baseline interferometry (VLBI). The GR prediction is $\Delta\phi_{\text{GR}} = 4M_\odot/R_\odot \simeq 1.75'' \simeq8.484\times 10^{-6} \text{rad}$ \cite{Will:2014kxa}.
The measurements agree with GR to within about $0.02\%$ (see, e.g., Ref.~\cite{Will:2014kxa} for a review). Thus the allowed deviation is:
\begin{equation}
|\delta(\Delta\phi)_\omega| \lesssim 0.02\% \times 8.484 \times 10^{-6} \ \text{rad} = 1.696 \times 10^{-9} \ \text{rad}.
\end{equation}
Using Eq.~\eqref{eq:deflection_correct} with $r_0 = R_\odot \simeq 6.96 \times 10^8$ m, the Weyl correction is such that
\begin{equation}
\frac{2 \times 6.96 \times 10^8}{\omega} \lesssim 1.696 \times 10^{-9} \hspace{0.1cm} \Rightarrow\hspace{0.1cm}  \omega \gtrsim 8.21 \times 10^{17} \ \text{m},
\end{equation}
which, in geometric units, reads 
\begin{equation}
\frac{\omega}{M_\odot} \gtrsim  5.56 \times 10^{14}.
\end{equation}
Let us now analyze the Solution II. In this case, the first-order approximation does not produce any correction to the GR prediction. Therefore, it is necessary to use the exact solution given in Eq.~(\ref{sol2}).

Following the same procedure as before, the Eq. (\ref{eq:null_simplified}) takes the form 
\begin{equation}
	\left(\frac{\mathrm{d} u}{\mathrm{d}\phi}\right)^2=({u_0}^2-u^2)-2M({u_0}^3-u^3)+\frac{M}{2\omega^2}(u_0-u). 
\end{equation}
Differentiating both sides with respect to $\phi$ and assuming $\frac{\mathrm{d} u}{\mathrm{d}\phi}\neq0$, we obtain at the null orbit equation
\begin{equation}
	\frac{\mathrm{d}^2 u}{\mathrm{d}\phi^2} + u = 3M u^2 - \frac{M}{4\omega^2}.
	\label{eq:null_II}
\end{equation}

Proceeding as in the previous case and treating the exact solution as a perturbation of the straight-line trajectory, the total deflection angle is given by
\begin{equation}
	\Delta\phi = \frac{4M}{r_0} - \frac{M}{2\omega^2} r_0.
	\label{eq:deflection_II}
\end{equation}
The Weyl correction is 
\begin{equation}
\delta(\Delta\phi)_\omega= - \frac{M}{2\omega^2} r_0.
\end{equation}
Using the same observational bound and considering $r_0=R_\odot\simeq6.96\times 10^8\text{m}$ and $M=M_\odot\simeq 1.477\times 10^3\text{m}$,  the allowed deviation is  such that
\begin{equation}
\frac{M_\odot R_\odot}{2\omega^2} \lesssim 1.696 \times 10^{-11} \hspace{0.1cm} \Rightarrow\hspace{0.1cm}  \omega \gtrsim  1.74 \times 10^{10} \ \text{m},
\end{equation}
which, in geometric units, reads 
\begin{equation} 
\frac{\omega}{M_\odot} \gtrsim 1.18\times 10^7.
\end{equation}
The two solutions lead to corrections with the same sign for the light-deflection angle. In both cases, the Weyl contribution is negative and therefore reduces the total deflection predicted by GR. Consequently, the Weyl correction acts in opposition to the standard Schwarzschild light-deflection angle.

Solar System light-deflection measurements yield \(\frac{\omega}{M_\odot} \gtrsim 5.56\times10^{14}\) for Solution I, whereas for Solution II one finds \(\frac{\omega}{M_\odot} \gtrsim 1.18\times10^{7}\). The latter bound is therefore approximately eight orders of magnitude weaker than the former.

\section{Radar Echo Delay}
\label{sec:radarecho}

We now consider the radar echo delay. Since the propagation of radio signals is described by radial null geodesics, the corresponding time delay can be obtained from the null-geodesic equations. Considering ${\mathrm{d}\theta=\mathrm{d}\phi=0}$ and $\mathrm{d}s^2 = 0$ we have
\begin{equation}
-e^{\alpha(r)} \mathrm{d}t^2 + e^{\beta(r)} \mathrm{d}r^2 = 0 \hspace{0.1cm} \Rightarrow\hspace{0.1cm}  \frac{\mathrm{d}t}{\mathrm{d}r} = \pm \sqrt{\frac{e^{\beta(r)}}{e^{\alpha(r)}}}.
\end{equation}
Using $e^{\beta(r)} = e^{-\alpha(r)}$ (which holds approximately for Solution I), the previous result can be rewritten as
\begin{equation}
\frac{\mathrm{d} t}{\mathrm{d} r} = \pm \frac{1}{e^{\alpha(r)}}.\label{eq:dtdr}
\end{equation}

Considering Solution I, to first order in small quantities, we can expand
\begin{equation}
\frac{1}{e^{\alpha(r)}} = 1 + \frac{2M}{r} - \frac{2r}{\omega} + \mathcal{O}(\epsilon^2).
\end{equation}
Thus the coordinate time for light to travel from $r_1$ to $r_2$ ($r_2 > r_1$) is given by
\begin{equation}
	\Delta t = \int_{r_1}^{r_2} \left(1 + \frac{2M}{r} - \frac{2r}{\omega}\right) \mathrm{d}r.
	\label{eq:radar_integral}
\end{equation}
Integrating the previous expression, it is possible to obtain
\begin{equation}
	\Delta t = (r_2 - r_1) + 2M \ln\left(\frac{r_2}{r_1}\right) - \frac{1}{\omega} ({r_2}^2 - {r_1}^2).
	\label{eq:radar_delay}
\end{equation}

The first two terms are the standard Schwarzschild result (the second term is the Shapiro delay). The extra delay due to the Weyl term is
\begin{equation}
	\delta(\Delta t)_\omega = -\frac{1}{\omega}({r_2}^2 - {r_1}^2).
	\label{eq:radar_extra}
\end{equation}

The Cassini spacecraft experiment measured the Shapiro delay to an accuracy of about $10^{-5}$~\cite{Bertotti:2003rm}. For a radar signal from Earth (at $r_E \simeq 1$ AU $= 1.496 \times 10^{11}$ m) to Saturn (at $r_S \simeq 9.5$ AU $\simeq 1.42 \times 10^{12}$ m) and back, the signal passes close to the Sun at ${r = r_0}$ (closest approach). However, for simplicity, we can use the approximation $r_0 \ll r_E, r_S$.

For a one-way trip from $r_0$ to $r_E$, the extra delay beyond the GR prediction is
\begin{equation}
\delta(\Delta t)_{\text{one-way}} = -\frac{1}{\omega}(r_E^2 - r_0^2) \simeq -\frac{r_E^2}{\omega},
\end{equation}
where we have neglected $r_0^2$ since $r_0 \ll r_E$.

For the round trip (Earth $\to$ Sun $\to$ Saturn $\to$ Sun $\to$ Earth), the total extra delay is approximately
\begin{equation}
\delta(\Delta t)_{\text{round}} \simeq -\frac{2}{\omega}({r_E}^2 + {r_S}^2).
\end{equation}
Using the observational data described above, we obtain 
\begin{equation}
|\delta(\Delta t)_\text{round}| \simeq \frac{4.078\times 10^{24}}{\omega} \text{m}.
\end{equation}
In geometric units ($c=1$), temporal time is measured in meters. Converting to SI units requires multiplication by $1/c = 3.3356 \times 10^{-9} \text{sm}^{-1}$, yielding
\begin{equation}
|\delta(\Delta t)|_{\text{seconds}} \simeq \frac{1.36 \times 10^{16}}{\omega} \ \text{s}.
\end{equation}

The Cassini experiment measured the round-trip time delay with an accuracy such that the GR prediction was confirmed to within $\sim 2 \times 10^{-9}$ s (see Ref.~\cite{Bertotti:2003rm} for details). Requiring the additional delay to remain below this observational uncertainty yields
\begin{equation}
\frac{1.36 \times 10^{16}}{\omega} \lesssim 2 \times 10^{-9} \hspace{0.1cm} \Rightarrow\hspace{0.1cm}  \omega \gtrsim 6.80 \times 10^{24} \ \text{m},
\end{equation}
which, in geometric units, reads 
\begin{equation}
\frac{\omega}{M_\odot} \gtrsim  4.60 \times 10^{21}.
\end{equation}

\vspace{0.25cm}

Let us now analyze the Solution II. In this case, to first order in small quantities, we can expand
\begin{equation}
\frac{1}{e^{\alpha(r)}} = 1 + \frac{2M}{r} + \frac{r^2}{4\omega^2} + \mathcal{O}(\epsilon^2)
\end{equation}
Considering the first order approximation and integrating Eq~(\ref{eq:dtdr}), it is possible to obtain
\begin{equation}
\Delta t = (r_2 - r_1) + 2M\ln\left(\frac{r_2}{r_1}\right) + \frac{1}{12\omega^2}({r_2}^3 - {r_1}^3).
\end{equation}
The extra delay term is
\begin{equation}
	\delta(\Delta t)_\omega = \frac{1}{12\omega^2}({r_2}^3 - {r_1}^3).
	\label{eq:radar_extra_II}
\end{equation}
Using the same assumptions and observational data as in the previous analysis, the Weyl correction for a round trip Earth--Saturn is 
\begin{equation}
\delta(\Delta t)_{\text{round}} = \frac{1}{6\omega^2}({r_E}^3 + {r_S}^3) \simeq \frac{4.78 \times 10^{35}}{\omega^2} \ \text{m}.
\end{equation}
Converting to SI units, we obtain
\begin{equation}
\delta(\Delta t)_{\text{seconds}} \simeq \frac{1.59 \times 10^{27}}{\omega^2} \ \text{s}.
\end{equation}
Requiring the additional delay to remain below ${2 \times 10^{-9}}$, we have
\begin{equation}
\frac{1.59 \times 10^{27}}{\omega^2} \lesssim 2 \times 10^{-9} \hspace{0.1cm} \Rightarrow\hspace{0.1cm}  \omega \gtrsim 8.92 \times 10^{17} \ \text{m},
\end{equation}
which, in geometric units, reads 
\begin{equation}
\frac{\omega}{M_\odot} \gtrsim  6.04 \times 10^{14}.
\end{equation}

\vspace{0.25cm}

The two solutions lead to qualitatively different corrections to the radar echo delay. For Solution I, the Weyl correction has a negative sign and therefore reduces the total time delay predicted by GR. In contrast, the correction associated with Solution II is positive and contributes an additional delay, increasing the GR prediction.

Although the correction in Solution II scales as \(\frac{1}{\omega^2}\), the radar echo delay still provides a meaningful constraint, \(\frac{\omega}{M_\odot} \gtrsim 6.04\times10^{14}\). Nevertheless, this bound remains substantially weaker than that obtained for Solution I, \(\frac{\omega}{M_\odot} \gtrsim 4.60\times10^{21}\), differing by approximately seven orders of magnitude. \\

The lower bounds on the Weyl parameter $\omega$ derived from the four Solar System tests are summarized in Table~\ref{Table-Results}. The perihelion shift of Mercury yields the strongest constraint on $\omega$. The different magnitudes of the resulting bounds arise from the different observational uncertainties of the respective tests, which lead to different sensitivities to the Weyl parameter.

\begin{table*}[h]
	\centering
	\caption{Summary of the lower bounds on the Weyl constant $\omega$, expressed in meters and solar masses,obtained from the four classical  Solar System tests for Solutions I and II.}
	\label{Table-Results}
	\begin{tabular}{lc@{\hspace{0.5cm}}c@{\hspace{0.5cm}}c@{\hspace{0.7cm}}c}
		\hline
		\textbf{Test} & \textbf{Solution} & $\omega \gtrsim (\text{m})$ & $\omega / M_\odot \gtrsim$ & \textbf{Key scaling} \\
		\hline 
		Gravitational redshift & I & $7.02 \times 10^{18}$ & $4.75 \times 10^{15}$ & $\delta z_\omega \propto 1/\omega$ \\
		& II & $3.63 \times 10^{14}$ & $2.53 \times 10^{11}$ & $\delta z_\omega \propto 1/\omega^2$ \\
		\hline
		Perihelion shift (Mercury) & I & $1.12 \times 10^{30}$ & $7.60 \times 10^{26}$ & $\delta(\Delta\phi)_\omega \propto 1/\omega$ \\
		& II & $1.53 \times 10^{20}$ & $1.04 \times 10^{17}$ & $\delta(\Delta\phi)_\omega \propto 1/\omega^2$ \\
		\hline
		Light deflection (Sun) & I & $8.21 \times 10^{17}$ & $5.56\times10^{14}$ & $\delta(\Delta\phi)_\omega \propto 1/\omega$ \\
		& II & $1.74 \times 10^{10}$ & $1.18 \times 10^{7}$ & $\delta(\Delta\phi)_\omega \propto 1/\omega^2$ \\
		\hline
		Radar echo delay (Cassini) & I & $6.80 \times 10^{24}$ & $4.60 \times 10^{21}$ & $\delta (\Delta t)_\omega \propto 1/\omega$ \\
		& II & $8.92 \times 10^{17}$ & $6.04 \times 10^{14}$ & $\delta (\Delta t)_\omega \propto 1/\omega^2$ \\
		\hline
	\end{tabular}
\end{table*}

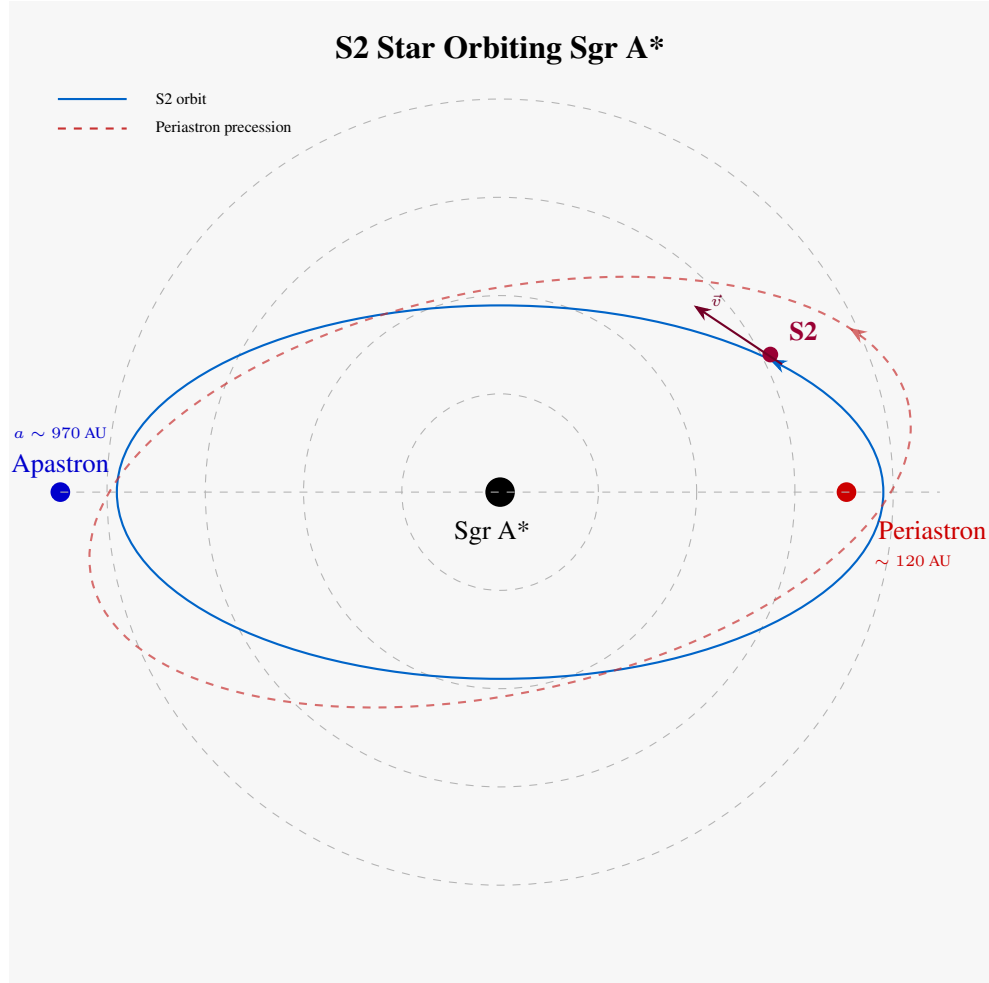
\begin{figure*}[htbp]
	\centering
	\begin{tikzpicture}[scale=1.3, >=Stealth]
		
		\definecolor{orbitcolor}{RGB}{0,100,200}
		\definecolor{precessioncolor}{RGB}{200,50,50}
		
		\fill[black!3] (-5,-5) rectangle (5,5);
		
		\foreach \r in {1,2,3,4} {
			\draw[black!30, dashed] (0,0) circle(\r);
		}
		
		\draw[orbitcolor, thick, 
		decoration={markings, mark=at position 0.1 with {\arrow{>}}}, 
		postaction={decorate}] 
		(0,0) ellipse [x radius=3.9, y radius=1.9];
		
		\fill[black] (0,0) circle(0.15);
		\node[below left, font=\small] at (0.4,-0.2) {Sgr A*};
		
		\fill[red!80!black] (3.525,0) circle(0.1);
		\node[below, font=\small, red!80!black] at (4.4,-0.2) {Periastron};
		\node[font=\tiny, red!80!black] at (4.2,-0.7) {$\sim 120$ AU};
		
		\fill[blue!80!black] (-4.475,0) circle(0.1);
		\node[above, font=\small, blue!80!black] at (-4.475,0.05) {Apastron};
		\node[font=\tiny, blue!80!black] at (-4.475,0.6) {$a\sim 970$ AU};
		
		\draw[black!30, dashed] (-4.475,0) -- (4.475,0);
		
		\draw[precessioncolor, thick, dashed, opacity=0.7,
		decoration={markings, mark=at position 0.05 with {\arrow{>}}},
		postaction={decorate}] 
		(0,0) ellipse [x radius=4.25, y radius=2.05, rotate=12];
		
		\fill[purple!80!black] (2.75,1.4) circle(0.08);
		\node[font=\small\bfseries, purple!80!black, above right] at (2.85,1.45) {S2};
		
		\draw[purple!60!black, ->, thick] (2.87,1.3) -- ++(-0.9,0.6);
		\node[font=\tiny, purple!60!black, above] at (2.2,1.8) {$\vec{v}$};

		\node[font=\large\bfseries] at (0,4.5) {S2 Star Orbiting Sgr A*};
		
		
		\draw[orbitcolor, thick] (-4.5,4.0) -- (-3.8,4.0);
		\node[font=\tiny, anchor=west] at (-3.6,4.0) {S2 orbit};
		\draw[precessioncolor, thick, dashed] (-4.5,3.7) -- (-3.8,3.7);
		\node[font=\tiny, anchor=west] at (-3.6,3.7) {Periastron precession};
		
	\end{tikzpicture}
	\caption{Schematic of the S2 star's orbit around Sgr A* at the Galactic Center. The blue ellipse shows the highly eccentric orbit (\(e \simeq 0.88\)) with semi-major axis \(a \simeq 970\) AU and pericenter distance of \(\sim 120\) AU. The red dashed ellipse illustrates the Schwarzschild periastron precession.}
	\label{fig:s2_orbit}
\end{figure*}

\section{Stellar Orbits Near Sgr A*}\label{Sgr}

The Galactic Center, hosting the supermassive black hole Sagittarius A* (Sgr A*), provides a unique laboratory for probing gravitational theories in the strong-field regime. The orbital motions of S-stars?particularly the S2 star with its 16-year period, high eccentricity ($e \simeq 0.88$), and pericenter distance of approximately 120~AU?have been monitored with extraordinary precision over the past three decades by the GRAVITY instrument at the Very Large Telescope and the Keck Observatory~\cite{GRAVITY:2018ofz,GRAVITY:2020gka,Do:2019txf}. A schematic representation of the S2 star orbiting Sgr A* is shown in Fig.~\ref{fig:s2_orbit}.
These observations have already detected relativistic effects including gravitational redshift and Schwarzschild precession at approximately $12'$ per orbit, consistent with GR at the $\sim 10\sigma$ confidence level, with the redshift parameter measured to approximately 10\% statistical accuracy \cite{GRAVITY:2018ofz}. The strong-field regime probed by these orbits ($\phi/c^2 \sim 10^{-3}$) is far more extreme than Solar System tests ($\phi/c^2 \sim 10^{-8}$), making them an ideal testing ground for the Weyl connection gravity solutions presented in this work.

In what follows, we analyze the implications of the two Schwarzschild-like solutions derived in Sec.~\ref{sec:BHsolutions} for stellar orbits near Sgr A*, deriving constraints on the Weyl parameter $\omega$ from current observations of the S2 star.

\vspace{0.1cm}

Considering  Solution I, the leading correction to the Schwarzschild metric scales as $\frac{1}{\omega}$, arising from the linear term $\frac{2r}{\omega}$. This linear correction produces observable effects in stellar orbits near Sgr A* that can be confronted with current data.

Following the same geodesic analysis as in Sec.~\ref{sec:perihelion}, the periastron precession per orbit for a test particle in Solution I is given by Eq.~(\ref{p}). The GR term $6\pi \frac{M}{p}$ is the standard Schwarzschild precession, while the Weyl correction enters with a negative sign, contributing a retrograde component that reduces the total relativistic precession.

For the S2 star orbiting Sgr A*, we adopt the parameters listed in the table \ref{tab:S2}. 
\begin{table}[htbp]
	\centering
	\caption{Orbital parameters of the S2 star near Sgr A*~\cite{GRAVITY:2020gka,Do:2019txf}.}
	\begin{tabular}{l|c}
		\hline
		Parameter & Value \\
		\hline
		Black hole mass & $M_\text{SgrA*} \simeq 4.0 \times 10^6 M_\odot \simeq 5.9 \times 10^9$ m \\
		Semi-major axis & $a \simeq 970$ AU $\simeq 1.45 \times 10^{14}$ m \\
		Eccentricity & $e \simeq 0.88$ \\
		Semi-latus rectum & $p = a(1-e^2) \simeq 219$ AU $\simeq 3.28 \times 10^{13}$ m \\
		Orbital period & $T \simeq 16$ yr \\
		\hline
	\end{tabular}
	\label{tab:S2}
\end{table}

The GR precession per orbit is:
\begin{equation}
	\begin{aligned}
	\Delta \phi_{\text{GR}} &= \frac{6\pi M_\text{SgrA*}}{p}  \simeq 3.39 \times 10^{-3} \text{ rad} \\
	&\simeq 0.194^\circ \simeq 12' \text{ per orbit}.
	\end{aligned}
\end{equation}
This matches the observed precession of S2~\cite{GRAVITY:2021xju}. The Weyl correction, given by Eq.(\ref{PerihelionCorrectionI}), is:
\begin{equation} \label{perihelion_SgrA}
	\delta(\Delta \phi)_\omega = -\frac{2\pi p^2}{\omega M_\text{SgrA*}}.
\end{equation}
The GRAVITY Collaboration has measured the precession parameter $f_{\text{SP}} = \Delta\phi / \Delta\phi_{\text{GR}} = 1.10 \pm 0.19$~\cite{GRAVITY:2020gka}. This corresponds to the interval $0.91\Delta\phi_{\text{GR}}\leq\Delta\phi\leq1.29 \Delta\phi_{\text{GR}}$, implying that any deviation from the GR prediction must satisfy  
\begin{equation}\label{deviationvalues}
-3.05\times 10^{-4}\text{rad}\leq\Delta\phi-\Delta\phi_{\text{GR}}\leq9.83\times 10^{-4}\text{rad}. 
\end{equation}
Since Eq.(\ref{perihelion_SgrA}) predicts a negative correction to the GR value, only the lower observational bound is relevant. Therefore, the Weyl correction must satisfy $\delta(\Delta\phi)_\omega \geq-3.05\times 10^{-4}\text{rad}$, which immediately yields
\begin{equation} 
\frac{1.15\times 10^8}{\omega}\lesssim3.05\times 10^{-4} \hspace{0.1cm} \Rightarrow\hspace{0.1cm}  \omega\gtrsim3.8\times 10^{21}~\text{m},
\end{equation}
which, in geometric units, reads
\begin{equation}
\frac{\omega}{M_\text{SgrA*}}\gtrsim 6.4\times 10^{11}. 
\end{equation}

The gravitational redshift for photons emitted from S2 near pericenter can also probe Solution I. The redshift correction scales as Eq.~(\ref{eq:extra_redshift_I}) where $r_1$ is the emission radius (near pericenter) and $r_2$ is the observer distance from Earth to S2. For S2, the pericenter distance is $r_p = a(1-e) \simeq 970 \times 0.12 \simeq 116$ AU $\simeq 1.74 \times 10^{13}$ m, while the Earth--S2 distance is approximately the distance to Sgr A*, $r_2 \simeq 2.45 \times 10^{20}$ m. Since $r_2 \gg r_1$, we have $\delta z_\omega \simeq r_2/\omega$.
\\
The GRAVITY Collaboration has measured the combined gravitational redshift and transverse Doppler effect for S2, corresponding to $z \simeq 200$ km/s$/c \simeq 6.67 \times 10^{-4}$, with an accuracy of approximately $10\%$ \footnote{The GRAVITY Collaboration has measured the combined gravitational redshift and relativistic transverse Doppler effect for the S2 star, parameterizing the post-Newtonian contribution by a factor $f$, where $f=0$ corresponds to the Newtonian limit and $f=1$ to the full general relativistic prediction. From data collected up to and including the pericenter passage in May 2018, the collaboration robustly detected the combined relativistic effect, corresponding to $z = \Delta\lambda/\lambda \simeq 200\ \text{km}\,\text{s}^{-1}/c \simeq 6.67 \times 10^{-4}$ \citep{GRAVITY:2018ofz,GRAVITY:2020gka}. From posterior fitting with different weighting schemes, they obtained
	\[
	f = 0.90 \pm 0.09\big|_{\text{stat}} \pm 0.15\big|_{\text{sys}},
	\]
	corresponding to a statistical uncertainty of approximately $10\%$ \citep{GRAVITY:2018ofz}.
	\\
	For the purpose of deriving conservative constraints on the Weyl parameter $\omega$ in this work, we adopt this $10\%$ uncertainty level. This choice ensures that the resulting bounds are robust and do not overstate the sensitivity of current observations to deviations from GR. Using this uncertainty, the allowed deviation in the redshift is
	\[
	|\delta z_\omega| \lesssim 0.1 \times z_{\text{obs}} \simeq 0.1 \times 6.67 \times 10^{-4} = 6.67 \times 10^{-5}.
	\]
}~\cite{GRAVITY:2018ofz,GRAVITY:2020gka}. Requiring the Weyl correction to remain below this observational uncertainty yields:
\begin{equation}
	\frac{2.45 \times 10^{20}}{\omega} \lesssim 6.67 \times 10^{-5}
	\hspace{0.1cm} \Rightarrow\hspace{0.1cm} 
	\omega \gtrsim 3.67 \times 10^{24}~\text{m},
\end{equation}
which, in geometric units, reads:
\begin{equation}
	\frac{\omega}{M_{\mathrm{SgrA^*}}} \gtrsim 6.22 \times 10^{14}.
\end{equation}
This bound, although stronger than the Solar System redshift constraint for Solution I, remains several orders of magnitude weaker than the Mercury perihelion constraint.

Let us now analyze the Solution II. In this case, the leading correction to the Schwarzschild metric scales as $\frac{1}{\omega^2}$, arising from the quadratic term $-\frac{r^2}{4\omega^2}$. The linear term is suppressed by an additional factor of $\frac{M}{\omega}$ and is negligible, as discussed in Sec.~\ref{B}.

Following the same procedure as in Sec.~\ref{sec:perihelion}, the Weyl correction, given by Eq.(\ref{eq:delta_precess_II_dom}), is
\begin{equation}
\delta(\Delta\phi)_\omega=\frac{3\pi p^3}{4\omega^2 M_\text{SgrA*}.}
\end{equation}
Since this relation predicts a positive sign, corresponding to an additional prograde contribution that enhances the GR precession, only the upper observational bound in Eq.~(\ref{deviationvalues}) is relevant. Therefore the Weyl correction must satisfy $\delta(\Delta\phi)_\omega \leq 9.83\times 10^{-4} \text{rad}$, which immediately yields 
\begin{equation} 
\frac{1.41\times 10^{31}}{\omega^2} \lesssim 9.83\times 10^{-4} \hspace{0.1cm} \Rightarrow\hspace{0.1cm}  \omega\gtrsim 1.2\times 10^{17}~\text{m}, 
\end{equation}
which, in geometric units, reads 
\begin{equation} 
\frac{\omega}{M_\text{SgrA*}} \gtrsim 2.0\times 10^{7}. 
\end{equation}
For Solution II, the gravitational redshift correction from Eq.~(\ref{z}) scales as:
\begin{equation}
	|\delta z_\omega| \simeq \frac{r_2^2}{8\omega^2}.
\end{equation}
Using the same observational constraint from the GRAVITY Collaboration,  we obtain
\begin{equation}
	\frac{(2.45 \times 10^{20})^2}{8\omega^2} \lesssim 6.67 \times 10^{-5}
	\hspace{0.1cm} \Rightarrow\hspace{0.1cm} 
	\omega \gtrsim 1.06 \times 10^{22}~\text{m},
\end{equation}
which, in geometric units, reads:
\begin{equation}
	\frac{\omega}{M_{\mathrm{SgrA^*}}} \gtrsim 1.80 \times 10^{12}.
\end{equation}
This bound is significantly stronger than the Solar System redshift constraint for Solution II, though still weaker than the Mercury perihelion constraint.

The two solutions lead to qualitatively different corrections to the periastron precession of the S2 star around Sgr A*. For Solution I, the Weyl correction enters with a negative sign and therefore opposes the standard GR precession, reducing the total relativistic advance. In contrast, the correction associated with Solution II is positive and contributes an additional prograde component, enhancing the GR prediction.

Although the correction in Solution II scales as $1/\omega^2$, the S2 orbital precession still provides a meaningful constraint, yielding $\omega/M_{\mathrm{SgrA^*}} \gtrsim 2.0 \times 10^7$. Nevertheless, this bound remains substantially weaker than that obtained for Solution I, $\omega/M_{\mathrm{SgrA^*}} \gtrsim 6.4 \times 10^{11}$, differing by approximately five orders of magnitude. The redshift constraints derived above provide complementary bounds, with $\omega/M_{\mathrm{SgrA^*}} \gtrsim 6.22 \times 10^{14}$ for Solution I and $\omega/M_{\mathrm{SgrA^*}} \gtrsim 1.80 \times 10^{12}$ for Solution II, further extending the observational test of both solutions to the strong-field regime.

\begin{table*}[htbp]
	\centering
	\caption{Summary of the lower bounds on the Weyl constant $\omega$ obtained from the S2 stellar orbit observations near Sgr A* for Solutions I and II. The redshift bounds are derived using the $10\%$ uncertainty from the GRAVITY Collaboration measurements.}
	\label{Table-S2}
	\begin{tabular}{lccccc}
		\hline
		\textbf{Test} & \textbf{Solution} & $\omega \gtrsim (\text{m})$ & $\omega / M_{\mathrm{SgrA^*}} \gtrsim$ & \textbf{Key scaling} \\
		\hline 
		Periastron precession & I & $3.80 \times 10^{21}$ & $6.44 \times 10^{11}$ & $\delta(\Delta\phi)_\omega \propto 1/\omega$ \\
		& II & $1.20 \times 10^{17}$ & $2.03 \times 10^{7}$ & $\delta(\Delta\phi)_\omega \propto 1/\omega^2$ \\
		\hline
		Gravitational redshift & I & $3.67 \times 10^{24}$ & $6.22 \times 10^{14}$ & $\delta z_\omega \propto 1/\omega$ \\
		& II & $1.06 \times 10^{22}$ & $1.80 \times 10^{12}$ & $\delta z_\omega \propto 1/\omega^2$ \\
		\hline
	\end{tabular}
\end{table*}

\begin{figure*}[htbp]
	\centering
	\begin{tikzpicture}[scale=1.2, transform shape]
		\begin{axis}[
			xbar,
			xlabel={$\log_{10}\left(\frac{\omega}{M_\odot}\right)$},
			y dir=reverse,
			ytick={1,2,3,4,5,6,7,8,9,10,11,12,13,14},
			yticklabels={
				Gravitational redshift I, Perihelion shift (Mercury) I, Light deflection (Sun) I, Radar echo delay (Cassini) I, S2 redshift I, S2 periastron precession I,
				Gravitational redshift II, Perihelion shift (Mercury) II, Light deflection (Sun) II, Radar echo delay (Cassini) II, S2 redshift II, S2 periastron precession II
			},
			xmin=0, xmax=30,
			legend style={at={(0.98,0.98)}, anchor=north east, font=\small},
			bar width=6pt,
			yticklabel style={font=\footnotesize},
			]
			
			\addplot[fill=red!60, draw=red!80!black,bar shift=0pt] coordinates {
				(15.68,1)   
				(26.88,2)   
				(14.74,3)   
				(21.66,4)}; 
			\addplot[fill=red!25,draw=red!80!black,forget plot,bar shift=0pt,] coordinates{
				(24.56,5)   
				(18.41,6)   
			};
			
			\addplot[fill=blue!60, draw=blue!100!black,bar shift=0pt] coordinates {
				(11.40,7)   
				(17.02,8)   
				(7.07,9)    
				(13.78,10)}; 
			\addplot[fill=blue!25, draw=blue!100!black,forget plot,bar shift=0pt,] coordinates{
				(22.03,11)  
				(13.91,12)  
			};

		\end{axis}
	\end{tikzpicture}
	\caption{Comparison of the lower bounds on $\frac{\omega}{M_\odot}$ obtained from the four classical Solar System tests and the S2 stellar orbit around Sgr A*. Red bars correspond to Solution I (scales as $\frac{1}{\omega}$), while blue bars correspond to Solution II (scales as $\frac{1}{\omega^2}$). Darker shades denote constraints derived from Solar System tests, whereas lighter shades correspond to the S2 stellar-orbit constraints. The bars indicate the excluded values of $\frac{\omega}{M_\odot}$.}
	\label{fig:bounds}
\end{figure*}
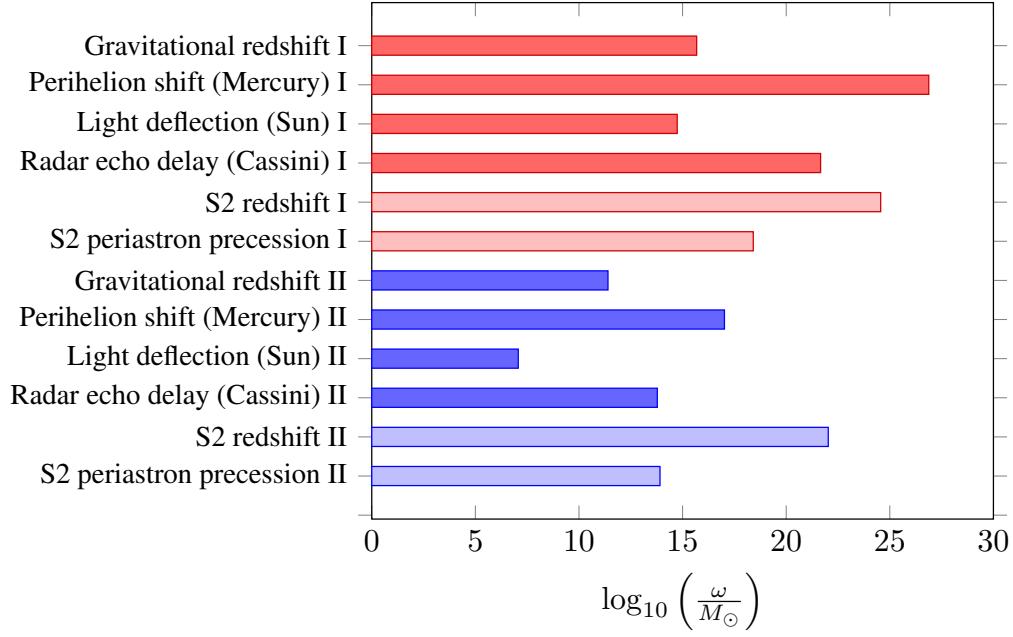

\section{Discussion and Conclusion}
\label{sec:conclusion}

In this work, we have analyzed the two Schwarzschild-like solutions found in Ref. \cite{Lima:2024cys}, and compared with both classical Solar System tests, and with the stellar orbits near galactic center. The constraints derived in the previous sections have an impact on the phenomenological viability of the Schwarzschild-like solutions in non-minimally coupled Weyl connection gravity. For convenience, the main results are summarized in Fig.~\ref{fig:bounds}, which collects the lower bounds on the Weyl constant $\omega$ obtained from the four classical Solar System tests and the S2 stellar orbit around Sgr A* (also have listed in Tables~\ref{Table-Results} and \ref{Table-S2} ). The observational constraints  reveal a clear distinction between the two Schwarzschild-like solutions. Although both are consistent with current Solar System tests and the S2 stellar orbit observations, the lower bounds obtained for the Weyl parameter differ by several orders of magnitude, reflecting the different scaling of the leading metric corrections with $\omega$.

Despite both solutions exhibiting quadratic corrections proportional to $\frac{r^2}{\omega^2}$ in the asymptotic limit, the behavior in the weak-field regime differs markedly. For Solution I, the linear term $\frac{2r}{\omega}$ dominates over the quadratic one for $r \ll \omega$, leading to observables that scale as $\frac{1}{\omega}$. In contrast, for Solution II, the linear term is suppressed by an additional factor of $\frac{M}{\omega}$, so that the quadratic term $-\frac{r^2}{4\omega^2}$ dominates also throughout the weak-field regime, yielding observables that scale as $\frac{1}{\omega^2}$. This distinction explains why the bounds on Solution I are consistently many orders of magnitude more stringent than those on Solution II.

For Solution I, the perihelion advance of Mercury provides the most restrictive constraint, requiring $\omega \gtrsim 10^{30}$ m. This scale is many orders of magnitude greater than the characteristic length scales probed within the Solar System, implying that the Weyl-induced corrections are effectively suppressed throughout the weak-field regime. Radar echo delay also yields a strong constraint ($\omega \gtrsim 10^{25}$ m), while the bounds derived from gravitational redshift and light deflection, although comparatively weaker ($\omega \gtrsim 10^{19}$ m and $\omega \gtrsim 10^{18}$ m, respectively), remain sufficiently restrictive to preclude observable departures from the Schwarzschild solution with current experimental precision.

In contrast, for Solution II, the perihelion shift and radar echo delay still require $\omega \gtrsim 10^{18}$--$10^{20}$ m, whereas light deflection yields a considerably less restrictive bound ($\omega \gtrsim 10^{10}$ m). This hierarchy reflects the reduced sensitivity of the light-deflection test to the higher-order Weyl corrections characterizing Solution II, suggesting that this observable may provide the best opportunity to probe the small deviations predicted by this solution as experimental precision continues to improve.

It is important to understand the physical origin of the significant differences in the lower bounds on \(\omega\) reported in Table~\ref{Table-Results}. These differences arise from two distinct sources. First, the scaling of the Weyl correction with \(\omega\) differs between the two solutions: for Solution~I the leading metric correction scales as \(1/\omega\), while for Solution~II it scales as \(1/\omega^2\), making the bounds for the latter systematically weaker by several orders of magnitude. Second, even within the same solution, each observable probes different combinations of the metric functions and their derivatives, leading to different sensitivities. In particular, the perihelion shift of Mercury accumulates over many orbits and is sensitive to the long-range behavior of the metric, yielding the most stringent constraint for both solutions (\(\omega \gtrsim 1.12 \times 10^{30}\,\mathrm{m}\) for Solution~I and \(\omega \gtrsim 1.53 \times 10^{20}\,\mathrm{m}\) for Solution~II). The radar echo delay, which involves large propagation distances and high-precision Cassini data, provides the second strongest bound, while the gravitational redshift and light deflection, being single-pass effects with comparatively lower sensitivity to the subleading corrections, yield the weakest constraints. The resulting hierarchy---perihelion shift \(\gg\) radar echo delay \(\gg\) gravitational redshift \(\gtrsim\) light deflection---is therefore a natural consequence of both the different \(\omega\)-scaling of the two solutions and the intrinsic sensitivity of each classical test to the Weyl-induced corrections.

The overall picture emerging from these results is that the weak-field regime provides a stringent consistency test for non-minimally coupled Weyl connection gravity. Rather than excluding the theory, the Solar System observations confine the Weyl parameter to a region where its effects are strongly suppressed, since all current Solar System bounds push $\omega$ to scales far exceeding the Solar System size ($\sim 10^{11}$ m). Consequently, any phenomenological distinction from GR is expected to become relevant only in regimes where the Weyl corrections are amplified, such as in strong gravitational fields or over cosmological distances.

From an observational perspective, these results suggest that detecting Weyl-induced corrections through the classical Solar System tests is unlikely with current experimental precision. Future improvements in precision (e.g., BepiColombo for Mercury perihelion, or next-generation VLBI for light deflection) may tighten the existing bounds, although substantial progress would be required before any deviation from GR could become observable within the weak-field regime.

It is also worth discussing the role of the parametrization used to describe test particle trajectories. Throughout the main analysis, the particle trajectories were parametrized by proper time, corresponding to the normalization $u^\mu u_\mu=-1$. In the presence of non-metricity, however, one may instead consider an affine parameter associated with the autoparallel transport defined by the generalized connection, for which the normalization becomes $u^\mu u_\mu=-\ell^2$. In this case, the proper time and the affine parameter cannot coincide, being related through $\mathrm{d} \tau=\pm\ell \mathrm{d} \lambda$. We therefore repeated the perihelion-shift analysis using this parametrization. For Solution I, the Weyl vector is sufficiently simple to allow an explicit determination of $\ell^2$, yielding, in the weak-field regime, $\ell^2\simeq1+2\frac{r}{\omega}$. Repeating the perihelion calculation with this modified normalization changes the resulting Weyl correction only by a factor of two. For Solution II, although the Weyl vector is not locally exact and an explicit expression for $\ell^2$ cannot be obtained directly from a path-independent integral, the angular component of the autoparallel transport equation implies that $\ell^2$ remains constant within the weak-field approximation considered. It can therefore be normalized to unity, recovering the proper time normalization. These results indicate that, within the weak-field regime relevant to the Solar System tests, the choice between proper time and affine parametrizations is not expected to lead to appreciable differences in the phenomenological bounds derived above.

The analysis of stellar orbits near Sgr A* provides an independent test in a substantially stronger gravitational regime than that of the Solar System. Despite probing a regime where $\phi/c^2 \sim 10^{-3}$, significantly more extreme than Solar System tests ($\phi/c^2 \sim 10^{-8}$), the current constraints from the S2 star are intermediate between the weakest and the strongest Solar System bounds. For Solution I, the S2 periastron precession yields 
$\omega \gtrsim 3.8 \times 10^{21}$~m, while the S2 gravitational redshift provides the stronger constraint 
$\omega \gtrsim 3.67 \times 10^{24}$~m. For Solution II, the S2 periastron precession yields $\omega \gtrsim 1.2 \times 10^{17}$~m, while the S2 gravitational redshift provides the stronger constraint 
$\omega \gtrsim 1.06 \times 10^{22}$~m. Although these bounds do not surpass the most stringent Solar System constraints, they are considerably stronger than those derived from gravitational redshift and light deflection, and are comparable to the intermediate range of Solar System limits. This demonstrates that stellar dynamics around Sgr~A* provides a complementary observational probe of the theory, extending the parameter constraints to a gravitational regime that is inaccessible to Solar System experiments.

Nevertheless, the stellar-orbit analysis also reveals a qualitative distinction between the two solutions. Solution I predicts a retrograde correction to the periastron precession, thereby reducing the total relativistic advance, whereas Solution II predicts a prograde correction, enhancing the GR prediction, in agreement with the behavior found in the Solar System analysis. Although the predicted deviations remain below the sensitivity of current observations, the opposite signatures provide a clear phenomenological distinction between the two branches and may become testable with future high-precision observations, such as those expected from GRAVITY+ and next-generation very long baseline interferometry (VLBI) facilities~\cite{GRAVITYPlus:2022,GRAVITYPlus:2024}. As the S2 star approaches its next pericenter passage, expected around 2034, the constraints on any deviation from GR are expected to improve, with the Shapiro effect potentially becoming detectable~\cite{Hyman:2023}. Together with ongoing and future Event Horizon Telescope (EHT) observations, these complementary probes will provide increasingly stringent tests of non-minimally coupled Weyl connection gravity across distinct gravitational regimes, extending and complementing the constraints obtained from precision Solar System experiments.

As discussed during the paper, the asymptotic structure of the two solutions reveals an intriguing connection with the Schwarzschild--(anti-)de Sitter family. Using the most stringent results obtained in this work, namely 
$\omega \gtrsim 1.12 \times 10^{30}$~m from Mercury perihelion for Solution I and 
$\omega \gtrsim 1.53 \times 10^{20}$~m from Mercury perihelion for Solution II (or 
$\omega \gtrsim 1.06 \times 10^{22}$~m from the S2 redshift for Solution II, which is weaker than the Mercury bound), the corresponding effective cosmological constants, defined via $\Lambda_{\text{eff}}^{(\text{I})} = -\frac{3}{\omega^2}$ and $\Lambda_{\text{eff}}^{(\text{II})} = \frac{3}{4\omega^2}$, satisfy
\[
|\Lambda_{\text{eff}}^{(\text{I})}| \lesssim 2.49 \times 10^{-60} \, \text{m}^{-2}  \text{ and } \Lambda_{\text{eff}}^{(\text{II})} \lesssim 3.20 \times 10^{-41} \, \text{m}^{-2}.
\]
For comparison, the observed cosmological constant is $\Lambda_{\text{obs}} = 1.1 \times 10^{-52} \, \text{m}^{-2}$~\cite{Planck:2018vyg}. Therefore, while both solutions are fully consistent with present weak-field observations, only Solution II remains compatible with an effective cosmological constant of the same order of magnitude as the observed one. This implies that the theory can naturally accommodate an asymptotic curvature compatible with cosmological observations while remaining consistent with all classical Solar System tests. For Solution I, by contrast, the effective cosmological constant is constrained to be approximately eight orders of magnitude smaller than $\Lambda_{\text{obs}}$, indicating that its quadratic asymptotic term cannot fully reproduce the observed cosmic acceleration.

This distinction carries implications for cosmology. Solution II naturally realizes a de Sitter-like asymptotic behavior, characterized by a positive effective cosmological constant. Such a geometry is of cosmological interest, as it could potentially account for the late-time accelerated expansion of the Universe without introducing an explicit cosmological constant in the action. The fact that current Solar System constraints allow $\omega$ to be as small as $\sim 10^{20}\,\text{m}$ means that $\Lambda_{\text{eff}}^{(\text{II})}$ could be as large as $\sim 10^{-41}\,\text{m}^{-2}$, well above the observed value, leaving room for the theory to reproduce cosmic acceleration. By contrast, the anti-de Sitter-like behavior of Solution I, with its negative effective cosmological constant, is less attractive from a cosmological perspective. Thus, Solution II appears as the more cosmologically promising branch of the theory: it passes all classical Solar System tests with weaker constraints, naturally accommodates a positive effective cosmological constant that could be of the same order as the observed value, and provides a geometric origin for cosmic acceleration without requiring an explicit vacuum energy contribution.

We emphasize that this correspondence should be regarded as an effective geometrical analogy rather than as an identification of the Weyl-induced term with the vacuum energy responsible for cosmic acceleration. Instead, it highlights that non-minimally coupled Weyl connection gravity naturally generates Schwarzschild--(anti-)de Sitter-like asymptotic geometries without introducing an explicit cosmological constant in the gravitational action.

\backmatter

\bmhead{Acknowledgements}

ML acknowledges support from Fundo Regional da Ci\^{e}ncia e Tecnologia and Azores Government through the Fellowship M3.1.a/F/031/2022, and from FCT/Portugal and the Recovery and Resilience Plan (PRR), through CAMGSD, projects UID/04459/2025 and UID/PRR/04459/2025, and the H2020-MSCA-2022-SE project EinsteinWaves, GA No.101131233.




\end{document}